\documentclass[iop,apj,twocolappendix,revtex4]{emulateapj}

\def\XXint#1#2#3{{\setbox0=\hbox{$#1{#2#3}{\int}$}
     \vcenter{\hbox{$#2#3$}}\kern-.5\wd0}}

\usepackage[dvips]{color}
\usepackage{graphicx}
\usepackage{mathrsfs}
\usepackage{color}

\begin{document}
\title{Anomalous CO$_2$ Ice Toward \object{HOPS-68}: A Tracer of Protostellar Feedback}

\author{Charles A.~Poteet\altaffilmark{1,6}, Klaus M.~Pontoppidan\altaffilmark{2}, S.~Thomas Megeath\altaffilmark{1}, Dan M.~Watson\altaffilmark{3},\\Karoliina Isokoski\altaffilmark{4}, Jon E.~Bjorkman\altaffilmark{1}, Patrick D.~Sheehan\altaffilmark{3,5}, and Harold Linnartz\altaffilmark{4}}

\altaffiltext{1}{Department of Physics and Astronomy, 
The University of Toledo, 2801 West Bancroft Street, Toledo, OH 43606, USA; charles.poteet@gmail.com}

\altaffiltext{2}{Space Telescope Science Institute, 3700 San Martin Drive, Baltimore, MD 21218, USA}

\altaffiltext{3}{Department of Physics and Astronomy, University of Rochester, 
Rochester, NY 14627, USA}

\altaffiltext{4}{Sackler Laboratory for Astrophysics, Leiden Observatory, Leiden University, P.~O.~Box 9513, 2300 RA Leiden, The Netherlands}

\altaffiltext{5}{Steward Observatory, University of Arizona, 933 North Cherry Avenue, Tucson, AZ 85721, USA}

\altaffiltext{6}{Current address: New York Center for Astrobiology, Rensselaer Polytechnic Institute, 110 Eighth Street, Troy, NY 12180, USA}

\slugcomment{Accepted to ApJ: 2013 February 15}

\shorttitle{Anomalous CO$_{2}$ Ice Toward \object{HOPS-68}}
\shortauthors{Poteet et al.}

\begin{abstract}

We report the detection of a unique CO$_{2}$ ice band toward the deeply embedded, low-mass protostar \object{HOPS-68}.  Our spectrum, obtained with the Infrared Spectrograph onboard the \emph{Spitzer Space Telescope}, reveals a 15.2 $\micron$ CO$_{2}$ ice bending mode profile that cannot modeled with the same ice structure typically found toward other protostars.  We develop a modified CO$_{2}$ ice profile decomposition, including the addition of new high-quality laboratory spectra of pure, crystalline CO$_{2}$ ice.  Using this model, we find that 87-92\% of the CO$_{2}$ is sequestered as spherical, CO$_{2}$-rich mantles, while typical interstellar ices show evidence of irregularly-shaped, hydrogen-rich mantles.  We propose that (1) the nearly complete absence of unprocessed ices along the line-of-sight is due to the flattened envelope structure of \object{HOPS-68}, which lacks cold absorbing material in its outer envelope, and possesses an extreme concentration of material within its inner (10 AU) envelope region and (2) an energetic event led to the evaporation of inner envelope ices, followed by cooling and re-condensation, explaining the sequestration of spherical, CO$_{2}$ ice mantles in a hydrogen-poor mixture.  The mechanism responsible for the sublimation could be either a transient accretion event or shocks in the interaction region between the protostellar outflow and envelope.  The proposed scenario is consistent with the rarity of the observed CO$_{2}$ ice profile, the formation of nearly pure CO$_{2}$ ice, and the production of spherical ice mantles.  \object{HOPS-68} may therefore provide a unique window into the protostellar feedback process, as outflows and heating shape the physical and chemical structure of protostellar envelopes and molecular clouds.

\end{abstract}

\keywords{astrochemistry --- circumstellar matter --- methods: laboratory  --- stars: formation --- stars: individual (\object{HOPS-68}, \object{FIR-2}) --- stars: protostars}

\section{Introduction \label{sec:intro}}
Stars form in cores deeply embedded within dense molecular clouds by gravitational collapse.  As an inseparable part of the star formation process, released gravitational energy from the central protostar heats the surrounding infalling envelope, and angular momentum is shedded via powerful bipolar outflows. These feedback mechanisms serve to disperse protostellar envelopes, halt further accretion, and sculpt the physical and chemical properties of star-forming regions, ultimately contributing to the evolution and destruction of their environment and the generation of turbulence \citep[e.g.,][]{McKee07}.  Understanding the feedback loop between protostars and molecular clouds is challenging; in part, because protostellar evolution is a stochastic process where violent accretion events may punctuate periods of more moderate accretion \citep{Hartmann96b, Dunham12, Fischer12}.  Hence, most observations of any given protostar only provide a single evolutionary snapshot, potentially leading to a biased interpretation.  As a remedy, observers have long sought tracers of the chemical and physical histories of protostars and their environments \citep{Charnley97, Arce07}.  One such tracer is CO$_2$ ice. 

Since its first detection by the \emph{Infrared Astronomical Satellite} \citep{dHendecourt89}, solid CO$_{2}$ has proven to be an abundant and ubiquitous constituent of the interstellar medium.  Observations of the 4.27 and 15.2 $\micron$ vibrational modes toward the Galactic molecular environments of both quiescent dark clouds \citep{Whittet98, Whittet07, Whittet09, Bergin05, Knez05} and circumstellar envelopes of low- and high-mass protostars \citep{Gerakines99, Nummelin01, Boogert04, Pontoppidan08, Zasowski09, Cook11} have revealed a solid CO$_{2}$ abundance of $\sim$15$-$40\% with respect to H$_{2}$O.  Chemical models indicate that gas-phase formation of CO$_2$ is highly inefficient at low temperature, and solid CO$_{2}$ is therefore widely assumed to be produced through chemical reactions on the surfaces of icy grain mantles \citep{Tielens82}.  Despite many experimental efforts to investigate the formation routes to solid CO$_{2}$, the precise chemical pathway is still debated \citep{Roser01, Ruffle01, Mennella04, Oba10, Noble11, Ioppolo11, Garrod11}.

The spectral profile of the CO$_{2}$ bending mode is a sensitive diagnostic of the local molecular environment.  In observations toward cold, quiescent molecular clouds, the profile is invariably characterized by a relatively broad, single-peaked, asymmetric absorption profile at 15.2 $\micron$ \citep{Knez05, Bergin05, Whittet07}.  The quiescent profile can generally be modeled as a linear combination of hydrogen-rich (H$_2$O:CO$_2$) and CO-rich (CO:CO$_2$) ice mixtures, typically dominated by the hydrogen-rich component, at low temperatures (\emph{T} $<$ 20 K).

At the other extreme, along lines-of-sight toward highly luminous (\emph{L} $\approx$ 10$^{4}\,$\emph{L}$_{\sun}$), massive protostars, the bending mode often exhibits a double-peaked substructure near 15.15 and 15.27 $\micron$ \citep{Gerakines99}. The double-peaked substructure is attributed to Davydov splitting, a phenomenon observed in molecular crystals having more than one equivalent molecule per unit cell \citep{Davydov62}, and is characteristic of pure, crystalline CO$_{2}$ \citep{Ehrenfreund97, Ehrenfreund99, vanBroekhuizen06}. The presence of pure CO$_2$ is generally interpreted as the result of thermal processing of icy grains seen along quiescent lines-of-sight.  \citet{Gerakines99} modeled the formation of pure CO$_2$ as a \emph{segregation} process of the hydrogen-rich component, requiring strong heating to high temperatures (\emph{T} $\approx$ 100 K, or close to the desorption temperature of the H$_{2}$O ice).  \citet{Pontoppidan08} suggested, by modeling the presence of CO$_{2}$ ice toward low-luminosity (\emph{L} $\approx$ 1$\,$\emph{L}$_{\sun}$) protostars, that pure CO$_2$ is also produced following the thermal desorption of CO from the CO:CO$_2$ mixture.  This \emph{distillation} process has the advantage that it can occur at much lower temperatures (\emph{T} = 20$-$30 K).  In either case, the band profile of solid CO$_2$ is an \emph{irreversible} tracer of thermal processing.  That is, by observing CO$_2$ ice, it can be determined whether the molecular cloud material has been heated significantly above $\sim$20$-$30$\,$ K, even if it has since re-cooled.  As such, past accretion outbursts have recently been invoked to explain the formation of pure CO$_{2}$ ice in low-luminosity (\emph{L} $\lesssim$ 1$\,$\emph{L}$_{\sun}$) protostars \citep{Kim12}. 

\citet{Pontoppidan08} found that all observed CO$_2$ bending mode profiles toward protostellar environments can be decomposed, phenomenologically, into five unique components.  Through the comparison with laboratory ice analogs, these components are ascribed to pure CO$_{2}$ ice, CO$_{2}$ mixed with CO or H$_{2}$O ice, dilute CO$_{2}$ mixed with pure CO ice, and CO$_{2}$ mixed with annealed CH$_{3}$OH and H$_{2}$O ices.  Each component has a fixed profile, but their relative column densities vary from source-to-source, likely reflecting variations in the pristine-to-processed ice fraction along the line-of-sight.  Moreover, this ``unique component'' structure appears to be a common property of other ice bands formed near protostars \citep[e.g., CO and the 6$-$8 $\micron$ complex;][]{Pontoppidan03, Boogert08}.  

There exists however, at least one exception to the unique component model; in this paper, we demonstrate that the 15.2 $\micron$ CO$_2$ ice band profile observed toward the peculiar protostar, \object{HOPS-68}, cannot be modeled with the same CO$_{2}$ ice structure that characterizes other protostars.  Originally identified as the compact millimeter source \object{FIR-2} \citep{Mezger90}, the object is located in the Orion Molecular Cloud 2 region \citep[OMC-2;][]{Gatley74} at an adopted distance of 414 $\pm$ 7 pc \citep{Menten07}.  \citet{Poteet11} showed that \object{HOPS-68} is a moderately luminous (1.3 \emph{L}$_{\sun}$) protostar with a flattened, relatively dense, infalling envelope that is viewed at an intermediate line-of-sight ($\theta \lesssim$ 41$\degr$).  From the observed spectral energy distribution (SED), they estimate a mass infall rate of \emph{\.{M}} = 7.6 $\times$ 10$^{-6}$ \emph{M}$_{\sun}$ yr$^{-1}$.  

\object{HOPS-68} is the first protostar to exhibit unambiguous evidence for the presence of crystalline silicates in \emph{absorption} \citep{Poteet11}, indicating that the silicates in its protostellar envelope have undergone strong thermal processing (\emph{T} $\gtrsim$ 1000 K).  While the mechanisms responsible for such processing are still not fully understood, \citet{Poteet11} proposed that amorphous silicates were annealed or vaporized within the warm inner region of the disk and/or envelope and subsequently transported outward by entrainment in protostellar outflows.  Alternatively, an \emph{in situ} formation by outflow-induced shocks may be responsible for the production of such material.  As such, \object{HOPS-68} might offer us a rare window into the effects of outflows on the solid component of protostellar envelopes. 

In Section \ref{sec:obs}, an overview of the spectroscopic observations, data reduction, and optical depth spectra are presented.  In Section \ref{sec:ana}, we demonstrate that the 15.2 $\micron$ bending mode profile toward \object{HOPS-68} is anomalous, and derive a new model for the CO$_2$ ice structure.  Finally, in Section \ref{sec:discus} we discuss possible physical scenarios that may explain the anomalous CO$_{2}$ ice structure toward \object{HOPS-68} in the context for more typical low-mass protostars.  Appendix \ref{sec:lab} presents new high-quality laboratory spectra of pure, crystalline CO$_{2}$ ice that are deployed in our analysis. 

\section{The \emph{Spitzer}-IRS Spectrum}\label{sec:obs}

\subsection{Observations and Data Reduction}
A high-resolution 10$-$37 $\micron$ spectrum of \object{HOPS-68} ($\alpha$~=~05$^{\rm{h}}35^{\rm{m}}24\fs30$, $\delta$~=~$-05\degr08\arcmin30\farcs6$ [J2000]) was obtained on 2008 November 14 using the \emph{Spitzer Space Telescope} \citep{Werner04} Infrared Spectrograph \citep[IRS;][]{Houck04}.  The observations (\emph{Spitzer} AOR 26614016) were performed with the high-resolution ($\lambda$/$\Delta\lambda$ $\sim$600) IRS modules, Short-High (SH; 9.9$-$19.6 $\micron$) and Long-High (LH; 18.7$-$37.2 $\micron$), at each of the two nominal nod positions, one-third of the way from the slit ends.  The total exposure times were 120 and 240$\,$s in the SH and LH modules, respectively, and were split between on- and off-source sky pointings.  

\begin{figure}[htp]
\centering
\rotatebox{90}{\includegraphics*[width=0.589\textwidth]{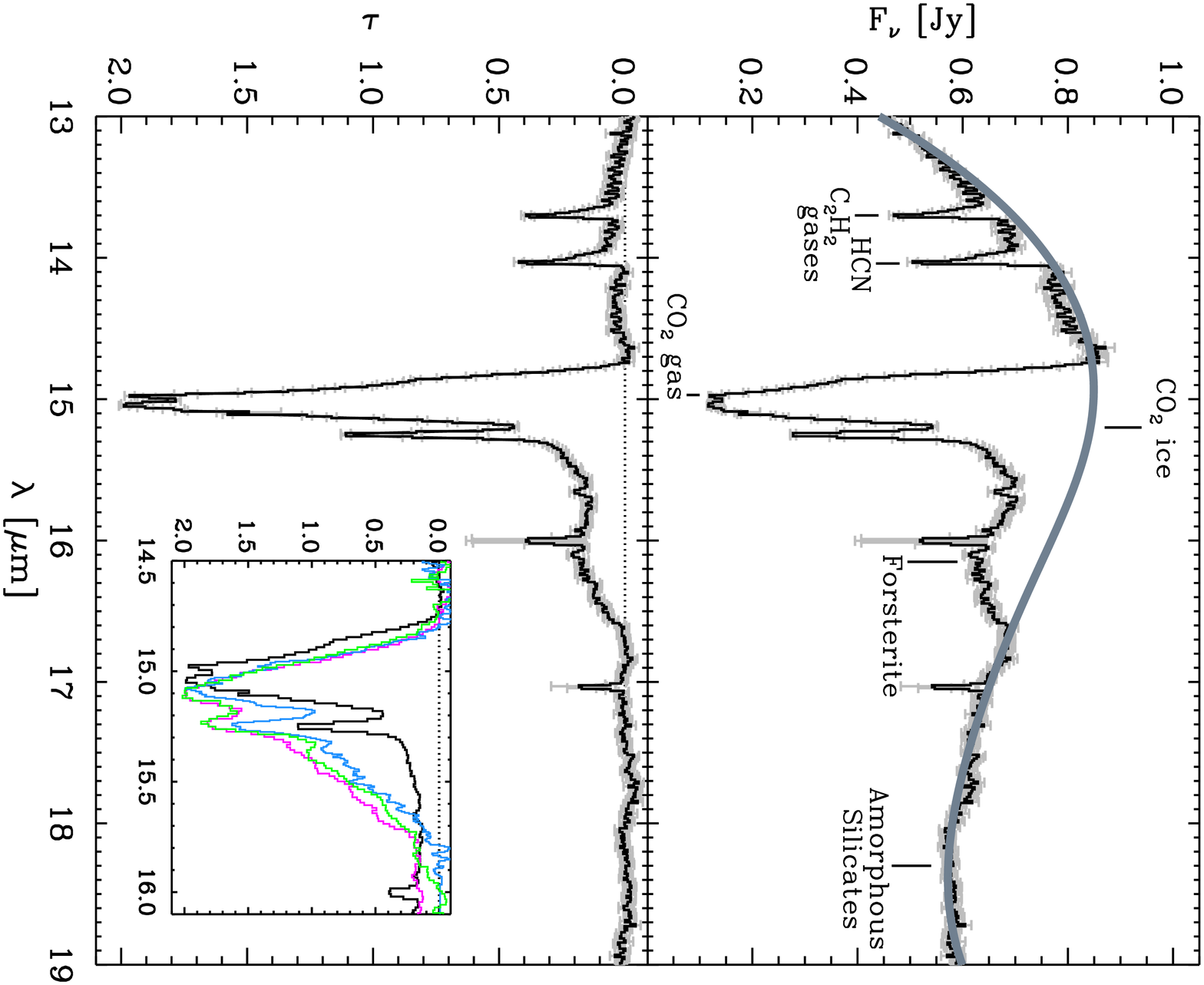}}
\caption{Top panel: \emph{Spitzer}-IRS 13$-$19 $\micron$ SH spectrum of HOPS-68, shown with statistical uncertainties (gray lines) and the adopted continuum (thick gray line).  The broad absorption near 16.1 $\micron$ has been ascribed to forsterite \citep[Mg$_{2}$SiO$_{4}$;][]{Poteet11}, and the narrow lines at 16.0 and 17.03 $\micron$ are background-subtraction artifacts; the latter resulting from H$_{2}$ \emph{S}(1) emission near the source.  Bottom panel:  derived optical depth spectrum of HOPS-68 (black line), centered near the strong 15.2 $\micron$ CO$_{2}$ ice $\tilde{\nu}_{2}$ bending mode.  The sharp absorption feature at 14.97 $\micron$ corresponds to the blended \emph{Q}-branch ro-vibrational lines of gaseous CO$_{2}$.  Insert:  close-up of the 14.5$-$16 $\micron$ spectrum of HOPS-68 (black line), with scaled optical depth profiles from the protostars: S140 IRS 1 (blue line; $\times$9.0), RNO 91 (green line; $\times$3.4), and IRAS 03254 (magenta line; $\times$4.5) shown for comparison.}
\label{fig:irs}
\end{figure}

The spectrum was extracted from the \emph{Spitzer} Science Center (SSC) S18.7 pipeline basic calibrated data using the SMART software package \citep{Higdon04}.  Permanently bad and `rogue' pixels were identified and corrected for by interpolation over nearby good pixels in the dispersion direction of the two-dimensional image.  In order to remove sky emission from the high-resolution data we averaged the sky observations and subtracted their mean from the on-source data for \object{HOPS-68}.  The sky subtracted two-dimensional images were averaged by nod position, and extracted using a full slit extraction.  Spectra of \object{Markarian 231} (LH) and \object{$\alpha$ Lac} (SH) were produced in the same manner, and relative spectral response functions (RSRFs) were produced by dividing a template by these spectra.  Each nod of the \object{HOPS-68} spectrum was then multiplied by these RSRFs to calibrate the flux density scale and the nods were averaged to obtain a final spectrum.  Finally, the SH and LH spectra were scaled to the flux density of the second-order of the Long-Low spectrum (LL2; 14.2$-$20.4 $\micron$) from \citet{Poteet11}.  The resultant SH spectrum is shown in the top panel of Figure \ref{fig:irs}. 

\subsection{Optical Depth Spectrum}

We convert the CO$_2$ flux density to an optical depth spectrum, using: $\tau(\lambda)$ = $-\ln$($F_{\lambda}^{\rm{obs}}$/$F_{\lambda}^{\rm{cont}}$), where $F_{\lambda}^{\rm{obs}}$ and $F_{\lambda}^{\rm{cont}}$ are the observed and continuum flux densities, respectively.  The continuum is constructed by fitting a third-order polynomial to the observed flux density within the wavelength ranges of 13.0$-$13.3, 14.6$-$14.7, and 18.2$-$19.5 $\micron$; the 13.5$-$14.2 $\micron$ range is avoided due to the rare gas-phase absorption of C$_{2}$H$_{2}$ (13.71 $\micron$) and HCN (14.05 $\micron$).  The polynomial is then combined with a Gaussian in the wavenumber domain, with a center at $\tilde{\nu}$ = 608 cm$^{-1}$ and a Full Width at Half Maximum (\emph{FWHM}) of $\Delta \tilde{\nu}$ = 73 cm$^{-1}$, to simulate the blue wing of the 18 $\micron$ silicate bending mode \citep{Pontoppidan08}.  

\section{An Anomalous CO$_2$ Ice Band}\label{sec:ana}

The derived CO$_2$ ice optical depth spectrum of \object{HOPS-68} is presented in the bottom panel of Figure \ref{fig:irs}.  The profile exhibits prominent double-peaked substructure at 15.05 and 15.24 $\micron$, characteristic of pure, crystalline CO$_{2}$ ice \citep{Hudgins93, Ehrenfreund97}.  The CO$_{2}$ bending mode profiles toward the well-studied protostars: \object{S140 IRS 1}, \object{RNO 91}, and \object{IRAS 03254} are shown for comparison in the insert of Figure \ref{fig:irs}, illustrating that the \object{HOPS-68} profile is fundamentally different \citep{Pontoppidan08, Gerakines99}.  The double-peaked substructure toward these low- and high-mass protostars occurs near 15.1 and 15.23 $\micron$ and is less prominent in comparison; the dip-to-peak ratios, defined by \citet{Zasowski09} as the local minimum to local maximum ratio of the blue peak, is 1.6$-$3.4 times less than that of \object{HOPS-68}.  Secondly, toward these protostars a third peak or broad shoulder is often observed near 15.4 $\micron$, but is only weakly detected in the SH spectrum of \object{HOPS-68}.  Finally, the SH spectrum of \object{HOPS-68} reveals the presence of both gaseous 14.97 $\micron$ CO$_{2}$ absorption and a blue absorption component or wing shortward of 15.05 $\micron$.  In comparison, gas-phase CO$_{2}$ absorption is weakly observed in the spectrum of the massive protostar \object{S140 IRS 1}; however, the spectrum displays no evidence for a blue wing.   

\begin{figure*}[htp]
\centering
\includegraphics*[width=0.75\textwidth]{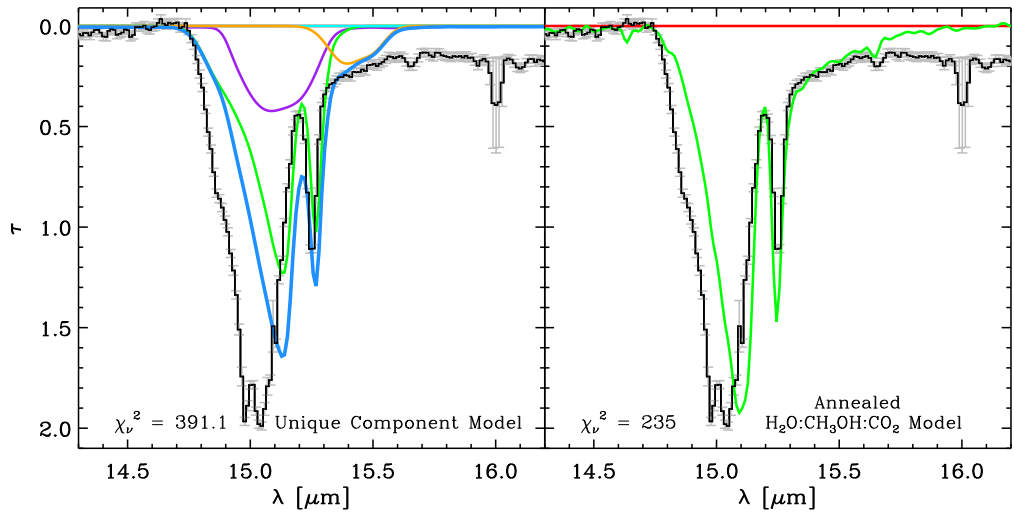}
\caption{\emph{Spitzer}-IRS SH optical depth spectrum of HOPS-68 (black line) and best-fit CO$_{2}$ ice models.  Left panel: the best-fit unique component CDE grain shape model (blue line) from \citet{Pontoppidan08}.  The model includes contributions from the pure CO$_{2}$ component (\emph{T} = 15 K; green line), the hydrogen-poor CO$_{2}$:CO = 112:100 mixture (\emph{T} = 10 K; violet line), and the 15.4 $\micron$ shoulder component (orange line).  Right panel:  the best-fit annealed ice model adopted from \citet{Gerakines99}.  The model consists of the annealed H$_{2}$O:CH$_{3}$OH:CO$_{2}$ = 1:0.9:1 mixture (\emph{T} = 125 K; green line) and the CDE shape-corrected laboratory spectrum of the hydrogen-rich H$_{2}$O:CO$_{2}$ = 100:14 mixture (\emph{T} = 10 K; red line).  The fitting procedures are constrained to wavelengths shortward of 15.6 $\micron$ to avoid the blue wing of the 16.1 $\micron$ forsterite feature.}
\label{fig:CO2models}
\end{figure*}

\subsection{Unique Component Model}\label{sec:cde}

We first demonstrate that the \emph{unique component decomposition}, which generally fits the CO$_2$ bending mode profiles in both low- and high-mass protostars, fails for \object{HOPS-68}.  Following \cite{Pontoppidan08}, we use the following five components, corrected for shape-effects using a continuous distribution of ellipsoids \citep[CDE;][]{Bohren83} model: hydrogen-rich CO$_{2}$ (H$_{2}$O:CO$_{2}$ = 100:14; \emph{T} = 10 K), CO-rich CO$_{2}$ (CO:CO$_{2}$ = 100:70, 100:26 or CO$_{2}$:CO = 112:100; \emph{T} = 10 K), pure CO$_{2}$ (\emph{T} = 15 K), and dilute CO$_{2}$ (CO:CO$_{2}$ = 100:4; \emph{T} = 10 K).  For the pure CO$_{2}$ component, we adopted the shape-corrected, high-resolution spectra from this work (see Appendix \ref{sec:lab}).  To match the resolving power of the SH module, the high-resolution spectra are convolved with a Gaussian, having a FWHM of $\Delta\lambda$ $\approx$ 0.03 $\micron$, at each wavelength element.  The fifth component, often referred to in the literature as the CO$_{2}$ shoulder at 15.4 $\micron$, has been identified as an intermolecular interaction between CO$_{2}$ and CH$_{3}$OH in annealed H$_{2}$O:CH$_{3}$OH:CO$_{2}$ mixtures \citep{Ehrenfreund98, Dartois99}, and is empirically modeled using a superposition of two Gaussians in the wavenumber domain.  

We employ a nonlinear least-squares fitting routine \citep[MPFIT;][]{Markwardt09} to determine the best-fit scaling of each component to the observed CO$_{2}$ bending mode of \object{HOPS-68}:
\begin{equation}
\tau(\lambda) = \sum_{i = 1}^5 \alpha_{i} \tau_{i}(\lambda),
\end{equation}
\noindent where $\tau_{i}(\lambda)$ is the normalized optical depth of an individual laboratory ice component and $\alpha_{i}$ is its corresponding scaling parameter.  The fit is performed over the wavelength ranges of 14.75$-$14.95 and 15.0$-$15.6 $\micron$ to avoid the gas-phase CO$_{2}$ absorption at 14.97 $\micron$ and the blue wing of the 16.1 $\micron$ forsterite feature.  

A comparison between the best-fit model and the observed CO$_{2}$ bending mode of \object{HOPS-68} is shown in the left panel of Figure \ref{fig:CO2models}, and clearly indicates that the ice composition is poorly simulated ($\chi_{\nu}^{2}$ $\approx$ 391) by the unique component decomposition from \citet{Pontoppidan08}.  The best-fit model results in a three-component ice mantle that is dominated by pure CO$_{2}$, with contributions from the CO-rich CO$_{2}$:CO = 112:100 mixture and the 15.4 $\micron$ shoulder component.  The hydrogen-rich H$_{2}$O:CO$_{2}$ = 100:14 mixture, which generally dominates the bending mode profiles toward low- and high-mass protostars \citep{Pontoppidan08, Gerakines99, Oliveira09, Seale11, An11, Kim12}, is entirely absent from the model.  The red wing of the observed profile is adequately matched by the pure CO$_{2}$ component, however the blue peak (15.14 $\micron$) of the pure CO$_{2}$ spectrum does not coincide with the observed peak position of 15.05 $\micron$.  Moreover, we find that the CO-rich CO$_{2}$:CO = 112:100 mixture is too broad and redshifted to account for the observed blue wing of the bending mode; decreasing the CO$_{2}$ fraction of this mixture results in a narrow profile that becomes further redshifted.    

\subsection{Annealed H$_{2}$O:CH$_{3}$OH:CO$_{2}$ Model}

We now consider the approach from \citet{Gerakines99}, who first interpreted the CO$_{2}$ bending mode profiles toward high-mass protostars using a combination of a hydrogen-rich mixture and an annealed CH$_{3}$OH-rich mixture.  Similarly to \citet{Seale11}, a two-component model was assembled using the CDE shape-corrected H$_{2}$O:CO$_{2}$ = 100:14 (\emph{T} = 10 K) mixture and an annealed H$_{2}$O:CH$_{3}$OH:CO$_{2}$ mixture from the large library of laboratory spectra described in \citet{White09}.  However, since the blue peak of the observed bending mode is positioned near 15.05 $\micron$, we consider only the bluest ($\tilde{\nu}$ $\gtrsim$ 662 cm$^{-1}$ or $\lambda$ $\lesssim$ 15.1 $\micron$) H$_{2}$O:CH$_{3}$OH:CO$_{2}$ spectra for simplicity.  The annealed mixtures are described in Table \ref{tbl:labspectra}.  For consistency with other studies \citep[e.g.,][]{Gerakines99, Cook11}, grain shape corrections were not applied to the absorption spectra of annealed mixtures; the segregation of CH$_{3}$OH and CO$_{2}$ in laboratory ice mixtures is thought to result in a highly inhomogeneous structure that is no longer well represented by thin films \citep{Gerakines99, Ehrenfreund99}.               

\begin{deluxetable}{lccc}
\centering
\tablecolumns{4}

\tablecaption{Summary of Laboratory Absorption Spectra}

\tablehead{
\colhead{Ice} &
\colhead{Mixture} &
\colhead{\emph{T}} &
\colhead{Resolution} \\
\colhead{Composition} &
\colhead{Ratio} &
\colhead{(K)} &
\colhead{(cm$^{-1}$)}\vspace{0.025 in} \\
\cline{1-4}\vspace{-0.06 in} \\
\multicolumn{4}{c}{Hydrogen-rich Ice \citep{Ehrenfreund97}}}

\startdata
H$_{2}$O:CO$_{2}$ \dotfill & 100:14 & 10 & 1 \vspace{-0.06 in} \\
\cutinhead{Hydrogen-poor Ices \citep{Ehrenfreund97}} 
CO:CO$_{2}$ \dotfill & 100:4 & 10 & 1 \\
CO:CO$_{2}$ \dotfill & 100:4 & 30 & 1 \\
CO:CO$_{2}$ \dotfill & 100:8 & 10 & 1 \\
CO:CO$_{2}$ \dotfill & 100:8 & 30 &  1 \\
CO:CO$_{2}$ \dotfill & 100:16 & 10 & 1  \\
CO:CO$_{2}$ \dotfill & 100:16 & 30 & 1  \\
CO:CO$_{2}$ \dotfill & 100:21 & 10 & 1  \\
CO:CO$_{2}$ \dotfill & 100:21 & 30 & 1  \\
CO:CO$_{2}$ \dotfill & 100:23 & 10 & 1  \\
CO:CO$_{2}$ \dotfill & 100:23 & 30 & 1  \\
CO:CO$_{2}$ \dotfill & 100:26 & 10 & 1  \\
CO:CO$_{2}$ \dotfill & 100:26 & 30 & 1  \\
CO:CO$_{2}$ \dotfill & 100:70 & 10 & 1  \\
CO:N$_{2}$:CO$_{2}$ $\;$\dotfill & 100:50:20 & 10 & 1  \\
CO:N$_{2}$:CO$_{2}$ $\;$\dotfill & 100:50:20 & 30 & 1  \\
CO:O$_{2}$:CO$_{2}$ $\;$\dotfill & 100:10:23 & 10 & 1  \\
CO:O$_{2}$:CO$_{2}$ $\;$\dotfill & 100:10:23 & 30 & 1  \\
CO:O$_{2}$:CO$_{2}$ $\;$\dotfill & 100:11:20 & 10 & 1  \\
CO:O$_{2}$:CO$_{2}$ $\;$\dotfill & 100:11:20 & 30 & 1  \\
CO:O$_{2}$:CO$_{2}$ $\;$\dotfill & 100:20:11 & 10 & 1  \\
CO:O$_{2}$:CO$_{2}$ $\;$\dotfill & 100:20:11 & 30 & 1  \\
CO:O$_{2}$:CO$_{2}$ $\;$\dotfill & 100:50:4 & 10 & 1 \\
CO:O$_{2}$:CO$_{2}$ $\;$\dotfill & 100:50:4 & 30 & 1 \\
CO:O$_{2}$:CO$_{2}$ $\;$\dotfill & 100:50:8 & 10 & 1 \\
CO:O$_{2}$:CO$_{2}$ $\;$\dotfill & 100:50:16 & 10 & 1  \\
CO:O$_{2}$:CO$_{2}$ $\;$\dotfill & 100:50:16 & 30 & 1  \\
CO:O$_{2}$:CO$_{2}$ $\;$\dotfill & 100:50:21 & 10 & 1  \\
CO:O$_{2}$:CO$_{2}$ $\;$\dotfill & 100:50:21 & 30 & 1  \\
CO:O$_{2}$:CO$_{2}$ $\;$\dotfill & 100:50:32 & 10 & 1  \\
CO:O$_{2}$:CO$_{2}$ $\;$\dotfill & 100:54:10 & 10 & 1  \\
CO:O$_{2}$:CO$_{2}$ $\;$\dotfill & 100:54:10 & 30 & 1  \\
CO:O$_{2}$:N$_{2}$:CO$_{2}$ $\,$\dotfill & 100:50:25:32 & 10 & 1 \\
CO:O$_{2}$:N$_{2}$:CO$_{2}$ $\,$\dotfill & 100:50:25:32 & 30 & 1 \\
CO:O$_{2}$:N$_{2}$:CO$_{2}$:H$_{2}$O \,.\,.\,.\, & 25:25:10:13:1& 10 & 1 \\
CO:O$_{2}$:N$_{2}$:CO$_{2}$:H$_{2}$O \dotfill & 50:35:15:3:1& 10 & 1 \\
CO:O$_{2}$:N$_{2}$:CO$_{2}$:H$_{2}$O \dotfill & 50:35:15:3:1& 30 & 1 \\
CO$_{2}$:CO \dotfill & 112:100:1 & 10 & 1 \\
CO$_{2}$:CO \dotfill & 112:100:1 & 45 & 1 \\
CO$_{2}$:H$_{2}$O \dotfill & 6:1 & 10 & 1 \\
CO$_{2}$:H$_{2}$O \dotfill & 6:1 & 42 & 1 \\
CO$_{2}$:H$_{2}$O \dotfill & 6:1 & 45 & 1 \\
CO$_{2}$:H$_{2}$O \dotfill & 6:1 & 50 & 1 \\
CO$_{2}$:H$_{2}$O \dotfill & 6:1 & 55 & 1 \\
CO$_{2}$:H$_{2}$O \dotfill & 6:1 & 75 & 1 \\
CO$_{2}$:H$_{2}$O \dotfill & 10:1 & 10 & 1 \\
CO$_{2}$:H$_{2}$O \dotfill & 10:1 & 80 & 1 \\
CO$_{2}$:H$_{2}$O \dotfill & 100:1 & 10 & 1 \\
CO$_{2}$:H$_{2}$O \dotfill & 100:1 & 30 & 1 \\
CO$_{2}$:O$_{2}$ $\;$\dotfill & 1:1 & 10 & 1 \\
\cutinhead{Pure CO$_{2}$ Ices (This Work)}
Pure CO$_{2}$ $\;$\dotfill & \nodata & 15 & \phantom{1.}0.1  \\
Pure CO$_{2}$ $\;$\dotfill & \nodata & 30 & \phantom{1.}0.1  \\
Pure CO$_{2}$ $\;$\dotfill & \nodata & 45 & \phantom{1.}0.1  \\
Pure CO$_{2}$ $\;$\dotfill & \nodata & 60 & \phantom{1.}0.1  \\
Pure CO$_{2}$ $\;$\dotfill & \nodata & 75 & \phantom{1.}0.1  \\
\cutinhead{Annealed Ices \citep{White09}} 
H$_{2}$O:CH$_{3}$OH:CO$_{2}$ $\,$\dotfill & 1:0.9:1 & 125 & 1  \\
H$_{2}$O:CH$_{3}$OH:CO$_{2}$ $\,$\dotfill & 1:0.9:1 & 130 & 1  \\
H$_{2}$O:CH$_{3}$OH:CO$_{2}$ $\,$\dotfill & 1:1.7:1 & 110 & 1  \\
H$_{2}$O:CH$_{3}$OH:CO$_{2}$ $\,$\dotfill & 1:1.7:1 & 115 & 1  \\
H$_{2}$O:CH$_{3}$OH:CO$_{2}$ $\,$\dotfill & 1:1.7:1 & 120 & 1 
\enddata
\label{tbl:labspectra}
\end{deluxetable}

The best-fit model is presented in the right panel of Figure \ref{fig:CO2models}.  In contrast to past studies \citep[e.g.,][]{Gerakines99, Seale11}, we find that combining the CDE shape-corrected hydrogen-rich H$_{2}$O:CO$_{2}$ = 100:14 (\emph{T} = 10 K) mixture with an annealed H$_{2}$O:CH$_{3}$OH:CO$_{2}$ = 1:0.9:1 (\emph{T} = 125 K) mixture yields a poor goodness of fit ($\chi_{\nu}^{2}$ $\approx$ 235).  In particular, we find that the hydrogen-rich mixture, which is known to consistently dominate the bending mode, is simply not utilized by the nonlinear least-squares fitting routine due to its broad redshifted profile.  Instead, the fitting routine attempts to constrain the ice composition with a single annealed mixture.  However, even for the bluest annealed mixtures, we find that the position of the blue peak does not coincide with the observed peak position of $\sim$15.05 $\micron$.  This result demonstrates that the observed CO$_{2}$ bending mode cannot be modeled by combining a hydrogen-rich mixture with an annealed H$_{2}$O:CH$_{3}$OH:CO$_{2}$ mixture, or by a single annealed H$_{2}$O:CH$_{3}$OH:CO$_{2}$ mixture.  Moreover, this is consistent with the fact that the unique profile decomposition is capable of simulating the same lines-of-sight as the annealed ice mixtures \citep{Pontoppidan08}.

\subsection{Multi-Component Homogeneous Spheres Model}\label{sec:sph}

\begin{figure*}[htp]
\centering
\includegraphics*[width=0.75\textwidth]{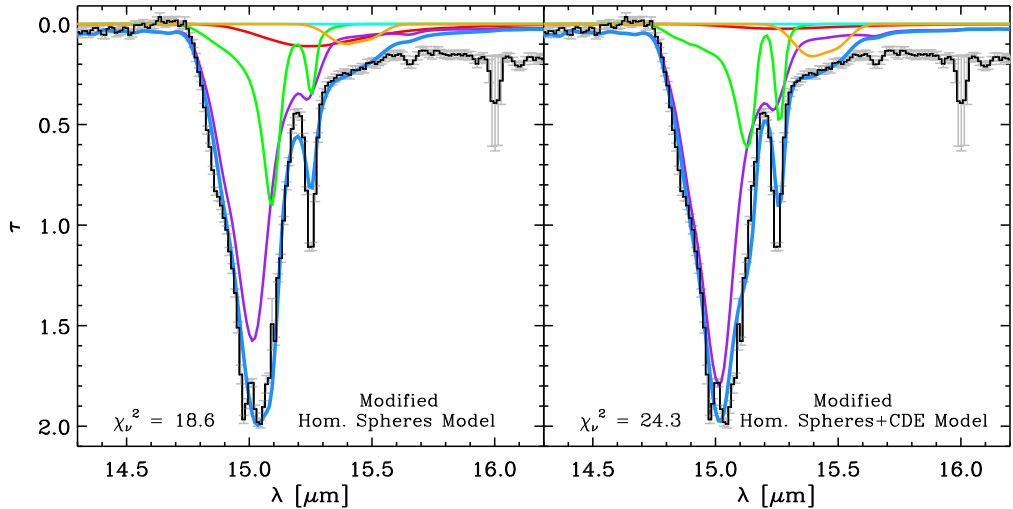}
\caption{\emph{Spitzer}-IRS SH optical depth spectrum of HOPS-68 (black line) and best-fit CO$_{2}$ ice models (blue lines).  Left panel: the best-fit multi-component homogeneous spheres model.  Individual CO$_{2}$ ice components include: hydrogen-poor CO$_{2}$ CO$_{2}$:H$_{2}$O = 6:1 mixture (\emph{T} = 55 K; violet line), pure CO$_{2}$ (\emph{T} = 15 K; green line), hydrogen-rich H$_{2}$O:CO$_{2}$ = 100:14 mixture (\emph{T} = 10 K; red line), and the 15.4 $\micron$ shoulder (orange line).  Right panel:  the best-fit multi-component CDE + spheres model.  A line-of-sight consisting of two grain shape populations was simulated by combining the CDE shape-corrected laboratory spectra of pure CO$_{2}$ ice (\emph{T} = 75 K; green line) and the hydrogen-rich H$_{2}$O:CO$_{2}$ = 100:14 mixture (\emph{T} = 10 K; red line), with the homogeneous CO$_{2}$:H$_{2}$O = 6:1 spheres mixture (\emph{T} = 55 K; violet line).}
\label{fig:newCO2models}
\end{figure*}

Having found that the traditional methods for analyzing the CO$_2$ ice band profile fail for \object{HOPS-68}, we now investigate a broader set of possibilities.  In particular, we must include a component that can provide additional absorption in the blue wing of the profile.  To this end, we consider additional CO$_{2}$-bearing, hydrogen-poor ice analogs from \citet{Ehrenfreund97} and explore the effect of grain shape by using homogeneous spheres in the Rayleigh limit.  The hydrogen-poor mixtures, which are summarized in Table \ref{tbl:labspectra}, consist of two or more molecular constituents and span a laboratory temperature range of \emph{T} = 10$-$80 K.  Grain shape-corrected spectra for a variety of particle shape distributions are presented by \citet{Ehrenfreund97}.  For relatively dilute CO$_{2}$ (e.g., H$_{2}$O:CO$_{2}$ = 100:14 and CO:CO$_{2}$ = 100:26), grain shape effects are generally weak, and only minor differences in the bending mode profile exist between the CDE and spherical grain shape corrections.  Conversely, for CO$_{2}$-rich mixtures (e.g., pure CO$_{2}$ and CO$_{2}$:H$_{2}$O = 6:1), spherical grain shape corrections often result in a narrow blueshifted profile that is qualitatively consistent with the observed blue wing. 

Following a similar methodology as before, a spectral decomposition of the observed bending mode was performed using the multi-component model from \citet{Pontoppidan08}.  However, for this model, the choice of the hydrogen-poor CO$_{2}$ component was not restricted to the three CO-rich mixtures used by \citet{Pontoppidan08}, but instead selected from the aforementioned suite of CO$_{2}$-bearing ice analogs by \citet{Ehrenfreund97}.  The laboratory temperature of the pure CO$_{2}$ component was also permitted to vary from \emph{T} = 15$-$75 K.  Furthermore, we assume that the line-of-sight consists exclusively of small homogeneous spheres.  

\begin{deluxetable*}{lccccccccc}[htp]
\centering
\tablewidth{0pt}
\tablecolumns{10}
\tablecaption{Column Densities and Abundances for Best-fit CO$_{2}$ Ice Models \label{tbl:col_dens}}

\tablehead{
\colhead{Grain Shape} &
\colhead{\emph{N}(CO$_{2}$)} &
\colhead{H-rich} &
\colhead{CO-rich} &
\colhead{Pure} &
\colhead{Shoulder} &
\colhead{Dilute} &
\colhead{H-poor} &
\colhead{\emph{T}$_{\rm{pure}}$} &
\colhead{$\chi_{\nu}^{2}$}\vspace{0.025 in} \\
\colhead{Model} &
\colhead{(10$^{17}$ cm$^{-2}$)} &
\colhead{(\%)} &
\colhead{(\%)} &
\colhead{(\%)} &
\colhead{(\%)} &
\colhead{(\%)} &
\colhead{(\%)} &
\colhead{(K)} &
\colhead{} \vspace{0.025 in}\\
\cline{1-10}\vspace{-0.06 in} \\
\multicolumn{10}{c}{Multi-component Models} }

\startdata
CDE $\:\,$\dotfill & 19.83 $\pm$ 0.49 &  \nodata & 27 $\pm$ 3 & 65 $\pm$ 3 & 8 $\pm$ 12 & \nodata & \nodata & 15 & 391.1 \\
Spheres $\:$ \dotfill & 24.66 $\pm$ 0.59 & 10 $\pm$ 13 & \nodata & 22 $\pm$ 3 & 3 $\pm$ 11 & \nodata & 65 $\pm$ 3 & 15 & \phantom{1}18.6 \\
CDE{\tiny$+$}Spheres $\,.\,.\,$ & 24.53 $\pm$ 0.59 & \phantom{1}2 $\pm$ 52 & \nodata & 18 $\pm$ 3 & 6 $\pm$ \phantom{1}8 & \nodata & 74 $\pm$ 3 & 75 & \phantom{1}24.3 \\
\cutinhead{Median Abundance Toward Low-mass Protostars \citep{Pontoppidan08}\tablenotemark{a}}
CDE $\:\,$\dotfill & 8.46$^{16.20}_{\phantom{1}4.34}$ & 70$^{74}_{65}$ & 20$^{27}_{15}$ & 6$^{12}_{\phantom{1}1}$ & 3$^{5}_{2}$ & $<$0.1$^{1}_{0}$ & \nodata & 15 & \nodata
\enddata
\tablecomments{Uncertainties are based on statistical errors in the \emph{Spitzer}-SH spectrum and systematic errors in the continuum determination.  Grain shape models from this work include: irregular grains (CDE), pure homogeneous spheres (spheres), and two separate grain shape populations (CDE+Spheres).  For the CDE and spheres models, all CO$_{2}$ components consist of irregularly-shaped particles and homogeneous spheres, respectively.  For the CDE+spheres model, all components are representative of irregularly-shaped particles, except  for the hydrogen-poor CO$_{2}$ component which is representative of homogeneous spheres.} 
\tablenotetext{a}{Median abundances include upper and lower quartile values.}
\end{deluxetable*}

The best-fit homogeneous spheres model is presented in the left panel of Figure \ref{fig:newCO2models}.  A four-component model, consisting largely of pure CO$_{2}$ (\emph{T} = 15 K) and the hydrogen-poor CO$_{2}$:H$_{2}$O = 6:1 (\emph{T} = 55 K) mixture, was found to provide the best fit ($\chi_{\nu}^{2}$ $\approx$ 19) to the observed bending mode.  As qualitatively expected, we find that the observed blue wing is adequately simulated by spherical grain shape-corrected laboratory spectra of CO$_{2}$-rich mixtures.  Furthermore, the observed red wing is well matched by a combination of pure CO$_{2}$, the hydrogen-rich H$_{2}$O:CO$_{2}$ = 100:14 (\emph{T} = 10 K) mixture, and the 15.4 $\micron$ shoulder component.  However, due to the relatively low amplitude of the pure CO$_{2}$ red peak, the optical depth of the 15.24 $\micron$ peak is slightly under-predicted by the best-fit model.  Nonetheless, we conclude that an exclusive homogenous spheres model, consisting predominately of the hydrogen-poor CO$_{2}$:H$_{2}$O = 6:1 (\emph{T} = 55 K) mixture, yields an overall excellent fit to the observed CO$_{2}$ bending mode profile. 

\subsection{Homogeneous Spheres + CDE Model}\label{sec:sph+cde}

The grain shapes of CO and CO$_{2}$ ice mantles in the circumstellar envelopes of protostars have been constrained using multiple ice bands of the same species \citep{Boogert02, Pontoppidan03, Gerakines99}.  A simultaneous decomposition of these bands indicates that their observed profiles are consistent with those of small irregularly-shaped ice mantles, simulated by CDE grain shape models.  For this reason, invoking a model consisting exclusively of homogeneous spheres may seem unjustified.  However, we have demonstrated that utilizing a CDE grain shape model to simulate the CO$_{2}$ bending mode of \object{HOPS-68} is largely unsatisfactory.  Even when the choice of the hydrogen-poor component is not constrained to the CO-rich CO:CO$_{2}$ mixtures used by \cite{Pontoppidan08}, we find that the goodness of fit is generally poor ($\chi_{\nu}^{2}$ $\approx$ 90$-$680).  Although our analysis is restricted to laboratory experiments for which grain shape corrections have been applied, this result implies that the observed blue wing is inconsistent with all available CDE shape-corrected hydrogen-poor ice analogs.  In contrast, we have shown that small homogeneous spheres of the hydrogen-poor CO$_{2}$:H$_{2}$O = 6:1 (\emph{T} = 55 K) ice mixture can adequately fit the blue wing of the observed bending mode.

To further examine the robustness of this result, we next consider a modeled line-of-sight that is populated by both spherical and irregularly-shaped grains.  Following a similar approach as in Section \ref{sec:sph}, a CO$_{2}$ ice model was constructed by combining the previously utilized CDE shape-corrected laboratory spectra (hydrogen-rich CO$_{2}$, pure CO$_{2}$, and dilute CO$_{2}$), a spherical grain shape-corrected hydrogen-poor laboratory spectrum, and the empirical 15.4 $\micron$ CO$_{2}$ shoulder component.  The best-fit model is shown in the right panel of Figure \ref{fig:newCO2models}.  Indeed, we find that a four-component model, composed largely of the same homogeneous CO$_{2}$:H$_{2}$O = 6:1 (\emph{T} = 55 K) spheres mixture from Section \ref{sec:sph}, is required for a satisfactory fit ($\chi_{v}^{2}$ $\approx$ 24).  The red wing of the observed bending mode is well matched by the combination of the CDE shape-corrected spectra of pure CO$_{2}$ (\emph{T} = 75 K) and the hydrogen-rich H$_{2}$O:CO$_{2}$ = 100:14 (\emph{T} = 10 K) mixture, with the 15.4 $\micron$ shoulder component.  In comparison to the model results from Section \ref{sec:sph}, we conclude that a modeled line-of-sight composed of two separate grain shape populations (spherical and irregularly-shaped particles) provides an equally good fit.  This result demonstrates that the CO$_{2}$ bending mode profile toward \object{HOPS-68} is largely governed by the grain shape and chemical composition of the hydrogen-poor CO$_{2}$ component.  
 
\subsection{CO$_{2}$ Ice Column Density}

The column density along the line-of-sight for each best-fit CO$_{2}$ component was estimated from Equation \ref{eq:col}, adopting the pure CO$_{2}$ bending mode band strength of \emph{A} = 1.1 $\times$ 10$^{-17}$ cm molecule$^{-1}$ from \citet{Gerakines95}.  The optical depth spectra were integrated over the wavenumber range of 680.3$-$613.5 cm$^{-1}$ ($\lambda$ = 14.7$-$16.3 $\micron$) to include the long wavelength wing of the observed profile.  

To estimate the systematic contribution from the baseline subtraction, we first computed the standard error, $\sigma$, between the best-fit continuum and the SH spectrum.  The best-fit continuum was then offset by $\pm$1 $\sigma$ to produce new optical depth spectra and column density estimates.  To quantify the errors due to the baseline uncertainty, we then calculated the mean absolute difference between the best-fit column densities and the column densities obtained by offsetting the best-fit continuum.  These errors were then added in quadrature with the best-fit errors, which are returned by the least-squares minimization technique, to generate the final CO$_{2}$ ice column density uncertainties. 

The column densities and relative abundances for the best-fit CO$_{2}$ ice models are tabulated in Table \ref{tbl:col_dens}.  The modified spectral decomposition analysis gives us the following three observables: 1) Among the available laboratory ice analogs, homogeneous spheres of the hydrogen-poor CO$_{2}$:H$_{2}$O = 6:1 (\emph{T} = 55 K) mixture are required to match the blue wing of the observed CO$_{2}$ bending mode.  This component constitutes 65$-$74\% of the total CO$_{2}$ ice column density along the line-of-sight.  2) The red wing of the observed profile can be modeled by combining pure CO$_{2}$ ice and the 15.4 $\micron$ CO$_{2}$ shoulder component.  The pure CO$_{2}$ component constitutes 18$-$22\% of the total CO$_{2}$ ice column density, while only a minor (3$-$6\%) contribution is attributed to the 15.4 $\micron$ CO$_{2}$ shoulder component.  3) Only 2$-$10\% of the total CO$_{2}$ ice column density is composed of the hydrogen-rich component, while no evidence for the CO-rich component is found within the observed CO$_{2}$ bending mode.  These fundamental differences of \object{HOPS-68}, as compared to the ice structure toward typical low-mass protostars, are illustrated in Table \ref{tbl:col_dens}. 

To derive the CO$_{2}$ ice abundance, the H$_{2}$O ice column density toward \object{HOPS-68} was estimated from the 12 $\micron$ libration mode of the amorphous silicate-subtracted optical depth spectrum from \citet{Poteet11}.  Adopting an intrinsic band strength of \emph{A} =  2.9 $\times$ 10$^{-17}$ cm molecule$^{-1}$ for crystalline H$_{2}$O ice \cite[\emph{T} = 140 K;][]{Mastrapa09}, we calculate a H$_{2}$O ice column density of \emph{N}(H$_{2}$O) = 88.2 $\pm$ 0.7 $\times$ 10$^{17}$ cm$^{-2}$.  For the best-fit models from Sections \ref{sec:sph} and \ref{sec:sph+cde}, we calculate an average CO$_{2}$ to H$_{2}$O ice column density ratio of \emph{N}(CO$_{2}$)/\emph{N}(H$_{2}$O) = 0.28 $\pm$ 0.03.  This is consistent with abundances measured toward low-mass protostars \cite[\emph{N}(CO$_{2}$)/\emph{N}(H$_{2}$O) = 0.32 $\pm$ 0.02;][]{Pontoppidan08}, but is significantly greater than those observed for high-mass protostars \cite[\emph{N}(CO$_{2}$)/\emph{N}(H$_{2}$O) = 0.17 $\pm$ 0.03;][]{Gerakines99}.

\section{Discussion and Conclusions}\label{sec:discus}

\subsection{The Rarity of Highly Processed Ice}

The \emph{Spitzer}-IRS SH spectrum toward the low-mass protostar \object{HOPS-68} reveals a rare 15.2 $\micron$ CO$_{2}$ bending mode profile, exhibiting a prominent double-peaked substructure (near 15.05 and 15.24 $\micron$) and a strong blue absorption wing.  Our analysis demonstrates that the observed bending mode cannot be modeled using the unique component models from \citet{Pontoppidan08} or by combining a hydrogen-rich mixture with an annealed H$_{2}$O:CH$_{3}$OH:CO$_{2}$ mixture \citep{Gerakines99}.  However, when the \citet{Pontoppidan08} models are modified to include spherical grains and a broader selection of hydrogen-poor ice mixtures, an overall excellent fit to the observed CO$_{2}$ bending mode is obtained.

Our results indicate that most (87$-$92\%) of the CO$_{2}$ along the line-of-sight is present in ices where CO$_{2}$ is the dominant constituent.  In contrast, the majority of the CO$_{2}$ ice observed toward most low- and high-mass protostars is found in environments dominated by H$_{2}$O and CO.  For example, toward low-mass protostars, \citet{Pontoppidan08} showed that generally $\sim$70\% of the total CO$_{2}$ ice column density is composed of the hydrogen-rich component, while $\sim$20\% is attributed to the CO-rich component.  Furthermore, the median abundance of pure CO$_{2}$ ice to the total CO$_{2}$ ice column density is found to be $\sim$6\%.  

The nearly complete lack of unprocessed ice mantles along the line-of-sight toward \object{HOPS-68} is unexpected.  For example, toward the prominent low-mass sources \object{IRAS 03254} and \object{RNO 91}, comparable fractions of pure CO$_{2}$ (17$-$23\%) are reported by \citet{Pontoppidan08}.  However, a substantial fraction (70$-$75\%) of their total CO$_{2}$ ice column density is attributed to unprocessed ice mixtures (CO:CO$_{2}$ and H$_{2}$O:CO$_{2}$), which lie outside the distillation and segregation radii \citep{Oberg11}.  In this context, the ice mantles toward \object{HOPS-68} are uniquely different from those observed toward other protostars, indicating that the observed icy grains have undergone a higher degree of thermal processing.  

\subsection{The Origin of Processed Ice Toward HOPS-68}

Nearly 100 CO$_2$ ice spectra have been described in the literature, and the observed profile toward \object{HOPS-68} is unique among them.  It is therefore unlikely to find more than a few percent of \object{HOPS-68} ice analogs in a larger sample of protostars.  Is this rarity due to \object{HOPS-68} being fundamentally different than other protostars, or does the observed line-of-sight reveal processed ices that are present in other protostars, but which are typically obscured by their cold outer envelope?  

\subsubsection{The Lack of Unprocessed Ice}

A peculiar spectral property of \object{HOPS-68} is the combination of a relatively flat 8$-$70 $\micron$ SED and strong silicate absorption at 9.7 $\micron$ \citep[][Figure 3]{Poteet11}.  To reproduce these observed characteristics, \citet{Poteet11} employed a radiative transfer model based on the collapse of an isothermal sheet initially in hydrostatic equilibrium \citep{Hartmann94, Hartmann96a}.  They conclude that \object{HOPS-68} possesses a highly flattened ($\eta$ = 2.0) protostellar envelope, as described by the degree of asphericity ($\eta$ = $R_{\rm{max}}/H$, where \emph{R$_{\rm{max}}$} is the outer envelope radius and \emph{H} is the scale height of the initial sheet).  In comparison, \citet{Furlan08} found that most Taurus protostars do not require an initially flattened density distribution, but instead can be modeled by the collapse of an initially spherically symmetric cloud core \citep{Terebey84}.  Unlike spherical envelopes, an implication of the highly flattened structure is that cold material in the outer envelope is not expected to be present along the observed line-of-sight.  Indeed, we estimate that $\lesssim$15\% of the gas has temperatures below \emph{T} = 30 K.  Thus, the absence of the volatile CO-rich CO$_{2}$ component and the low abundance of the hydrogen-poor CO$_{2}$ component may be explained by the lack of cold, unprocessed dust along the line-of-sight.  This result suggests that highly processed ices may be present in the inner regions of other protostellar envelopes, but are likely obscured or diluted by the large, cold envelopes of unprocessed material typically surrounding them.

Furthermore, to simultaneously simulate the strong silicate absorption, the \citet{Poteet11} model requires infalling material to be deposited within 0.5 AU of the central protostar, as set by the centrifugal radius ($R_{c}$).  In contrast, $R_{c}$ values from 10 to 300 AU, with a median of 60 AU, were found for Taurus protostars \citep{Furlan08}.  For a centrifugal radius of $R_{c}$ = 0.5 AU, we estimate that more than 50\% of the infalling material along the line-of-sight is concentrated within 10 AU of the central protostar, where the potential for thermal processing of icy grains by radiation and outflow-induced shocks is the strongest.

Resolved images of protostellar envelopes show a wide range of envelope morphologies, from highly flattened to relative spheroidal structures, as well as a great deal of complexity \citep{Tobin10, Tobin11, Tobin12}.  The diversity of envelope morphologies may be due to formation of protostars in turbulent and often filamentary molecular clouds \citep[e.g.,][]{Andre10, Molinari10}.  Although the envelope of \object{HOPS-68} has not yet been resolved, SED modeling requires a highly flattened envelope which may be uncommon within the diversity of envelope morphologies typically found in molecular clouds.  This flattened envelope, when observed from an intermediate inclination, results in a line-of-sight that is less obscured by unprocessed material in the cold, outer envelope.

\subsubsection{The Formation of Spherical CO$_{2}$-rich Mantles}

Although a flattened envelope structure may explain the absence of cold primordial ices seen toward other protostars, it does not directly explain why a large fraction of the CO$_{2}$ ice seen toward \object{HOPS-68} is sequestered as spherical, CO$_{2}$-rich mantles, while typical interstellar ices show evidence of irregularly-shaped, hydrogen-rich mantles.  That is, in the absence of unprocessed ices along the line-of-sight, the observed CO$_{2}$ ice profile should be consistent with irregularly-shaped CO$_{2}$-rich ice mantles.  Irregularly-shaped ice mantles are generally understood to be the product of a slow aggregation of icy grain monomers in dense molecular clouds prior to protostellar collapse \citep{Ossenkopf93, Ormel09}.  In the case of \object{HOPS-68}, we propose that the spherical ice mantles may have formed as the volatiles rapidly freeze out in dense gas, following an energetic, but transient event that sublimated any primordial icy grain population within its inner envelope region.

Radiative processing by episodic accretion events have recently been proposed by \citet{Kim12} to be a viable mechanism for producing pure CO$_{2}$ ice, through the distillation process, toward low-luminosity protostars (\emph{L} $\lesssim$ 1$\,$\emph{L}$_{\sun}$).  Using the unique component decomposition from \citet{Pontoppidan08}, \citet{Kim12} report a median pure CO$_{2}$ ice abundance of 15\% toward six protostars; this value is six times less than the processed ice abundance found toward \object{HOPS-68}.  The detection of pure CO$_{2}$ ice toward low-luminosity protostars indicates that a transient phase of higher luminosity must have existed some time in their past \citep{Kim12}.  For the case of \object{HOPS-68}, we propose that heating beyond the sublimation temperature of H$_{2}$O ice \citep[\emph{T} $\approx$ 110 K;][]{Fraser01} will evaporate ice mantles close to the central protostar during the high-luminosity phase, resulting in the separation of icy grain conglomerates.  As the protostar later returns to its quiescent, low-luminosity phase, gas temperatures in the envelope decrease and H$_{2}$O and CO$_{2}$ rapidly ($t$ $<$ 10$^{3}$ yr) condense onto grain surfaces in the order of their condensation temperature \citep{Ossenkopf93}, provided that the gas is sufficiently dense ($n_{\rm{H_{2}}}$ $>$ $10^{4}$ cm$^{-3}$).  This mechanism naturally allows for the formation of nearly pure CO$_{2}$ ice mantles sequestered on spherical grains.  However, this scenario requires that the accretion outburst must have occurred recently to avoid the reproduction of irregular grains by renewed grain aggregation.  

Evidence for transient heating is often presumed from the variable nature of the accretion luminosity.  Photometric monitoring of \object{HOPS-68}, with the \emph{Herschel Space Observatory}, reveal a $\sim$21\% variation in the 70 $\micron$ flux over a two week period; this variability is most likely driven by variations in the mass accretion rate onto the central protostar \citep{Billot12}.  While larger increases in luminosity, over longer timescales, are certainly needed to heat dust grains to the sublimation temperature of H$_{2}$O ice, the observed variability indicates a rapidly changing accretion rate for \object{HOPS-68}, and hence, supports the possibility of a recent ($t$ $\ll$ $10^{6}$ yr) accretion outburst that raised the luminosity significantly above the current observed value.  

\begin{figure}[htp]
\centering
\rotatebox{90}{\includegraphics*[width=0.589\textwidth]{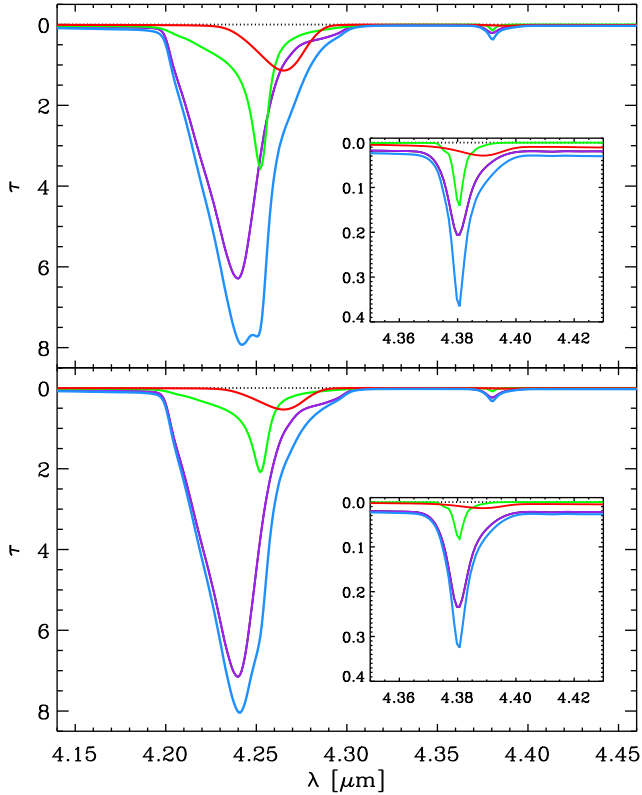}}
\caption{Model predictions for the ($\tilde{\nu}_{3}$) $^{12}$CO$_{2}$ and ($\tilde{\nu}_{3}$) $^{13}$CO$_{2}$ stretching modes at 4.27 and 4.38 $\micron$, respectively.  The predicted CO$_{2}$ profiles (blue lines) are constructed from the best-fit multi-component homogeneous spheres model (Section \ref{sec:sph}; top panel) and the multi-component CDE + homogeneous spheres model (Section \ref{sec:sph+cde}; bottom panel).  The individual ice components include: hydrogen-rich CO$_{2}$ (H$_{2}$O:CO$_{2}$ = 100:14 at \emph{T} = 10 K; red lines), hydrogen-poor CO$_{2}$ (CO$_{2}$:H$_{2}$O = 6:1 at \emph{T} = 55 K; violet lines), and pure CO$_{2}$ (\emph{T} = 15 and 75 K; green lines).}
\label{fig:stretchmod}
\end{figure}

In order to quantify the luminosity rise needed to produced the proposed sublimation of icy grains, we increase the total luminosity of the \citet{Poteet11} SED model by a factor of 10, 50, and 100.  Adopting a CO$_{2}$ ice desorption temperature of \emph{T} = 45 K, from an amorphous H$_{2}$O ice surface \citep{Noble12}, we assume that CO$_{2}$ ice is present in the envelope of \object{HOPS-68} at distances greater than \emph{r} $\approx$ 90 AU.  These incremental steps in luminosity result in increased dust temperatures of \emph{T} = 70, 97, and 113 K at \emph{r} $\approx$ 90 AU, respectively.  Thus, for \object{HOPS-68}, we propose that a factor of 100 increase in luminosity ($\sim$100 \emph{L}$_{\sun}$) is required to thermally desorb CO$_{2}$ and H$_{2}$O ices from the surface of dust grains.  Although the required luminosity is higher than the typical luminosities of protostars in Orion \citep[$\sim$1 \emph{L}$_{\sun}$;][]{Kryukova12}, luminosities in excess of 100 \emph{L}$_{\sun}$ are observed toward young, low-mass stars undergoing FU Orions-like outbursts \citep{Reipurth10}.  \citet{Kryukova12} finds five Orion protostars with luminosities exceeding 100 \emph{L}$_{\sun}$, including two candidate FU Orionis-like objects.  Furthermore, models developed to explain the typical low luminosities of protostars require that most low-mass protostars undergo FU Orionis-like outbursts in order to accrete the required amount of mass during the protostellar lifetime \citep{Dunham12}.

Alternatively, mechanical processing by shocks, associated with the observed outflow \citep{Williams03}, will also sublimate ice mantles in the outflow working surface along the cavity wall.  In dense shocks, dust grain temperatures are raised by collisional heating with hot gas particles and by absorption of photons as the compressed gas cools, resulting in thermal desorption and sputtering of interstellar ices.  In particular, for a nondissociative shock with a velocity of $v_{s}$ = 10 km s$^{-1}$, dust grains are expected to be heated to temperatures on the order of 10$^{2}$ K \citep{Neufeld94}, which is more than sufficient to allow for a condensation sequence to occur behind the shock.  Therefore, similar to the accretion outburst scenario, we propose that nearly pure CO$_2$ ice mantles are formed in post-shocked gas, again in a condensation sequence, which the line-of-sight must be dominated by shock-processed material near the cavity wall.  For other orientations, the line-of-sight will be dominated by irregularly-shaped icy grain mantles.  Toward \object{HOPS-68}, evidence for the presence of shock-processing may be inferred from the detection of [\ion{S}{1}], [\ion{Fe}{2}], and [\ion{Si}{2}] forbidden emission at 25.25, 25.98, and 34.82 $\micron$, respectively (D.~M.~Watson et al.~2013, in preparation), as well as gaseous absorption from C$_{2}$H$_{2}$, HCN, and CO$_{2}$ at 13.71, 14.05, and 14.97 $\micron$, respectively.  

We note that the formation of interstellar ices in a post-shocked gas was originally proposed by \citet{Bergin98, Bergin99}.  However, in their scenario, CO$_{2}$ ice is formed by virtue of gas-phase reactions (CO + OH $\rightarrow$ CO$_{2}$ + H) and subsequent freeze-out.  The formation process is largely governed by the photodissociation rate of H$_{2}$O molecules by cosmic-ray induced photons, and is expected to occur over a timescale of \emph{t} $\approx$ 10$^{5}$ yr at the assumed density of $n_{\rm{H_{2}}}$ = $10^{5}$ cm$^{-3}$; however, the timescale may be shortened with higher gas densities or larger cosmic-ray ionization rates.  Nonetheless, this long timescale for CO$_{2}$ ice mantle formation would allow for the re-development of irregularly-shaped grains, through the aggregation of small regular particles.  In contrast, the detection of spherical icy grains implies that ice mantle formation would have occurred more recently, over much shorter timescales (\emph{t} $\sim$ $\mathcal{O}$(10$^{2}$) yr) and at higher densities ($n_{\rm{H_{2}}}$ $\approx$ $10^{7}$ cm$^{-3}$ at $r$ = 100 AU), by evaporation and re-condensation processes alone.  

In summary, we propose that the observed 15.2 $\micron$ CO$_{2}$ ice profile may result from the combination of two circumstances.  First, the nearly complete absence of unprocessed ices along the line-of-sight is due to the highly flattened envelope of \object{HOPS-68}, which lacks cold absorbing material in its outer envelope, and possesses a large fraction of material within its inner (10 AU) envelope region.  Second, an energetic event led to the evaporation of inner envelope ices, followed by cooling and re-condensation, explaining the sequestration of CO$_{2}$ in a hydrogen-poor mixture and the spherical shape of the icy grains.  Regardless of which energetic process is responsible for producing the spherical mantles, we note that both proposed mechanisms require fortuitous timing and orientations, and hence are consistent with the observed rarity of spherical icy grains toward protostars. 

The notion of ices forming in a cooling gas, following an energetic event, makes several predictions.  Water ice condensed at high temperatures will arrange itself into a crystalline matrix, predicting that the H$_{2}$O ice seen toward \object{HOPS-68} should be predominantly crystalline.  Evidence for the presence of crystalline H$_{2}$O ice was predicted, by \citet{Poteet11}, from the 12 $\micron$ libration mode.  This prediction can be further tested by observing the profile of the 3.08 $\micron$ H$_{2}$O stretching mode.  Carbon monoxide ice, observed at 4.67 $\micron$, should either be absent if the temperature of the absorbing material is still above 20\,K, or should show evidence of spherical grain shapes, in sharp contrast to that observed toward typical protostars \citep{Pontoppidan03}.  Finally, the $^{12}$CO$_{2}$ and $^{13}$CO$_{2}$ stretching modes at 4.27 and 4.38 $\micron$, respectively, should be consistent with spherical grains and a nearly pure CO$_2$ mixture.  Using the best-fit models from Sections \ref{sec:sph} and \ref{sec:sph+cde}, predictions for the $^{12}$CO$_{2}$ and $^{13}$CO$_{2}$ stretching modes are presented in Figure \ref{fig:stretchmod}.  While \object{HOPS-68} is too faint at 3$-$5 $\micron$ for current spectrometers, these tests will likely be within reach of the \emph{James Webb Space Telescope}.

\begin{figure*}[htp]
\centering
\rotatebox{90}{\includegraphics*[width=0.495\textwidth]{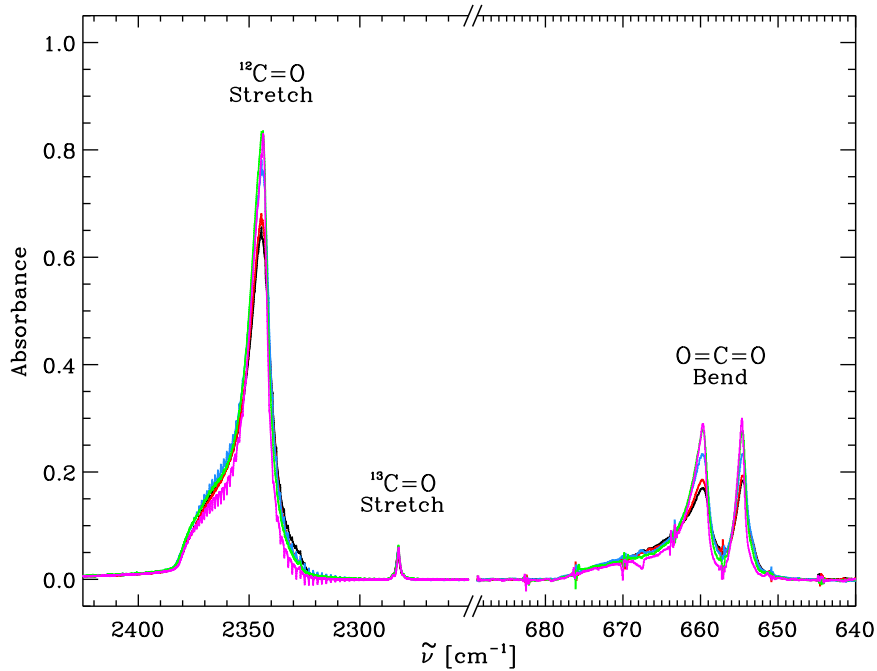}}
\caption{The background- and baseline-corrected absorption spectra of pure CO$_{2}$ ice.  The ($\tilde{\nu}_{3}$) $^{12}$CO$_{2}$ and ($\tilde{\nu}_{3}$) $^{13}$CO$_{2}$ stretching (2343 and 2283 cm$^{-1}$, respectively) and ($\tilde{\nu}_{2}$) $^{12}$CO$_2$ bending (660 and 655 cm$^{-1}$) modes are shown at laboratory temperatures of $T$ = 15 (black line), 30 (red line), 45 (blue line), 60 (green line), and 75 K (magenta line).  Gas-phase CO$_{2}$ at 668 cm$^{-1}$ is a background-subtraction artifact, arising from either under- or over-subtraction of background CO$_{2}$ gas.}
\label{fig:abs}
\end{figure*}

\begin{acknowledgements}
This work is based on observations made with the \emph{Spitzer Space Telescope}, which is operated by the Jet Propulsion Laboratory (JPL), California Institute of Technology (Caltech), under a contract with NASA.  Support for this work was provided by NASA through contracts 1289605 and 1355568 issued by JPL/Caltech.  This publication makes extensive use of data products from the Sackler Laboratory for Astrophysics at Leiden University and the Astrophysics Laboratory at the University of Alabama at Birmingham.  C.~A.~P wishes to thank Karin \"{O}berg for suggesting the laboratory experiments, and Pascale Ehrenfreund, Perry Gerakines, and Adwin Boogert for insightful discussions.
\end{acknowledgements}

\appendix

\section{Pure CO$_{2}$ Ice Laboratory Spectra}\label{sec:lab}

The infrared absorption bands of CO$_{2}$ have been extensively studied in the laboratory by \citet{Hudgins93}, \citet{Ehrenfreund97}, and \citet{Baratta98}.  Collectively, these studies present absorption spectra and optical constants for pure CO$_{2}$ at laboratory temperatures of \emph{T} = 10, 12, 30, 50, 70, and 80 K.  These experiments are conducted under high-vacuum conditions at a spectral resolution of 1$-$2 cm$^{-1}$, too low to fully resolve the narrow structure of the Davydov split.  More recently, CO$_{2}$ experiments with a spectral resolution of 0.5 cm$^{-1}$ have been conducted at laboratory temperatures of \emph{T} = 15$-$90 K \citep{vanBroekhuizen06}, but at relatively low signal-to-noise.

In this paper, we deploy new high-quality infrared absorption spectra and optical constants from a temperature-series laboratory study of pure, crystalline CO$_{2}$ ice.  The experiments were conducted at an improved spectral resolution of 0.1 cm$^{-1}$ and at higher signal-to-noise with respect to previous laboratory studies. 

\subsection{Experimental Procedure}

Following the procedures of \citet{Gerakines95} and \citet{Bouwman07}, the experiments are performed in a high vacuum (HV) chamber with a base pressure of \emph{P} $\approx$ 5$\times$10$^{-7}$ mbar at room temperature.  The chamber houses a CsI (Caesium Iodide) substrate that is cooled down to 15 K by a closed cycle helium cryostat (ADP DE-202).  The substrate temperature is controlled with a resistive heater element and a silicon diode sensor using an external control unit (LakeShore 330).  

A sample of CO$_{2}$ (Praxair, 99.998~$\%$ purity) is introduced into the system from a gas bulb at 10.0 mbar prepared in a separate vacuum manifold (base pressure of \emph{P} $\approx$ 10$^{-5}$ mbar).  Pure CO$_{2}$ ices are grown onto the substrate at 15 K via effusive dosing along the surface normal at a deposition rate of $\sim$10$^{15}$ molecules cm$^{-2}$ s$^{-1}$.  The ices are heated to temperatures of \emph{T} = 15, 30, 45, 60, and 75 K at a rate of 2 K min$^{-1}$ and allowed to relax at each temperature for five minutes before recording the absorption spectra.  A Fourier Transform InfraRed (FTIR) spectrometer (Varian 670-IR) is used to record the ice spectra in transmission mode over the wavenumber range of $\tilde{\nu}$ = 4000$-$400 cm$^{-1}$ ($\lambda$ = 2.5$-$25 $\micron$) with a spectral resolution of 0.1 cm$^{-1}$, averaging a total of 256 scans to increase the signal-to-noise ratio.  Background spectra are acquired at 15 K prior to deposition and subtracted from the recorded ice spectra.  

\subsection{Experimental Results}

\begin{figure}[htp]
\centering
\rotatebox{90}{\includegraphics*[width=0.387\textwidth]{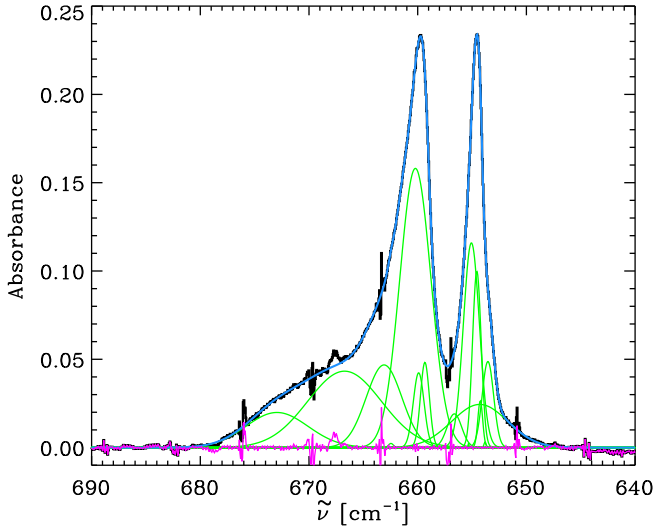}}
\caption{The $\tilde{\nu}_{2}$ bending mode spectrum of pure CO$_{2}$ ice (black line) compared to its corresponding ``smoothed" spectrum (blue line) at \emph{T} = 45 K.  A least-squares fit is performed using a superposition of Gaussians (green lines) to eliminate noise and gas-phase CO$_{2}$ absorption at 668 cm$^{-1}$.  The resultant fit residuals (magenta line) are shown for comparison.}
\label{fig:gauss}
\end{figure}

To facilitate a comparison with astronomical spectra, a baseline correction is applied to all laboratory absorption spectra by fitting a third-order polynomial over the wavenumber regions of 4000$-$3800, 2900$-$2700, 2270$-$2250, 950$-$700, and 645$-$640 cm$^{-1}$.  The background- and baseline-corrected absorption spectra for pure CO$_{2}$ ice are shown in Figure \ref{fig:abs} for the ($\tilde{\nu}_{3}$) $^{12}$CO and ($\tilde{\nu}_{3}$) $^{13}$CO stretching and ($\tilde{\nu}_{2}$) CO$_2$ bending vibrational modes.  

The spectra do not suffer from any significant contamination by background gases of H$_{2}$O within the vacuum system.  However, weak traces of gaseous CO$_{2}$, near 668 cm$^{-1}$ ($\lambda$ = 14.97 $\micron$), are present in some absorption spectra, and are the result from either an under- or over-subtraction of background CO$_{2}$ gas.  To reduce the noise and eliminate the gas-phase CO$_{2}$ absorption, a nonlinear least-squares minimization is simultaneously performed to the $\tilde{\nu}_{3}$ $^{12}$CO, $\tilde{\nu}_{3}$ $^{13}$CO, and $\tilde{\nu}_{2}$ CO$_2$ vibrational modes using a superposition of Gaussians.  Figure \ref{fig:gauss} illustrates the `smoothing' technique for the $\tilde{\nu}_{2}$ bending mode spectrum at $T$ = 45 K.  The measured peak positions ($\tilde{\nu}$) and widths (\emph{FWHM}; $\Delta\tilde{\nu}$) of the $\tilde{\nu}_{2}$ bending mode are listed in Table \ref{tbl:labpureco2}.  The position of the absorption peaks near 654.5 and 659.7 cm$^{-1}$ are temperature-independent and remain constant to within 0.1 cm$^{-1}$, while their \emph{FWHM}s decrease by a factor of two with increasing temperature from \emph{T} = 15 to 75 K.  These results are in agreement with the 0.5 cm$^{-1}$ resolution study by \citet{vanBroekhuizen06}.       

\subsection{Optical Constants}

The complex index of refraction (\emph{n} + \emph{ik}) is determined using a Kramers-Kronig analysis, following similar methods previously described in \citet{Hudgins93} and \citet{Mastrapa09}.  

Briefly, the imaginary part of the complex refractive index, \emph{k}, is calculated from the Beer-Lambert law:

\begin{equation}
\alpha(\tilde{\nu}) \equiv 4 \pi k \tilde{\nu} = \frac{1}{d} \ln \negthinspace\left(10\right) A_{\tilde{\nu}},
\end{equation}

\noindent where $\alpha$ is the Lambert absorption coefficient, \emph{d} is the thickness of the ice, and $A_{\tilde{\nu}}$ is the absorbance of the ice at wavenumber $\tilde{\nu}$.  The thickness of the ice samples were estimated using the known absorption band strength of pure CO$_{2}$ ice.  In this technique, the column density, \emph{N} (in molecules cm$^{-2}$), of the sample was determined using 

\begin{equation}\label{eq:col}
 N = \frac{\int \tau_{\tilde{\nu}} \, d\tilde{\nu}}{A}, 
\end{equation}

\noindent where $\tau_{\tilde{\nu}}$ = $\ln$(10)$ A_{\tilde{\nu}}$ is the optical depth and \emph{A} is the intrinsic band strength of the $\tilde{\nu}_{2}$ bending mode as determined from previous laboratory studies \citep[\emph{A} = 1.1$\times$10$^{-17}$ cm molecule$^{-1}$;][]{Gerakines95}.  The ice thickness is then inferred from the derived column density assuming a monolayer (ML) surface coverage of $\sim$10$^{15}$ molecules cm$^{-2}$.  The final ice thickness is tabulated in Table \ref{tbl:labpureco2} and ranges in value from 355$-$410 ML.  Assuming a monolayer thickness of 5.54 \AA, the lattice constant for solid CO$_{2}$ \citep{Keesom34}, these values correspond to a physical thickness of $\sim$0.20$-$0.23 $\micron$.    

\begin{deluxetable}{ccccc}
\centering
\tablewidth{0pt}
\tablecolumns{5}

\tablecaption{Band Positions and \emph{FWHM}s of the $\tilde{\nu}_{2}$ CO$_{2}$ Bending Mode}

\tablehead{
\colhead{\emph{T}} &
\colhead{$\tilde{\nu}$} &
\colhead{\emph{FWHM}, $\Delta \tilde{\nu}$} &
\multicolumn{2}{c}{Ice Thickness, \emph{d}} \\
\cline{4-5}
\colhead{(K)} &
\colhead{(cm$^{-1}$)} &
\colhead{(cm$^{-1}$)} &
\colhead{(ML)\tablenotemark{a}} & 
\colhead{($\micron$)}\vspace{0.025 in} \\
\cline{1-5}\vspace{-0.06 in} \\
\multicolumn{5}{c}{Laboratory Spectrum} }

\startdata
15 & 654.5/659.7 & 2.1/5.3 & 366 & 0.203 \\
30 & 654.5/659.8 & 2.0/4.3 & 367 & 0.203 \\
45 & 654.5/659.7 & 1.8/3.6 & 410 & 0.227 \\
60 & 654.6/659.7 & 1.5/2.9 & 405 & 0.224 \\
75 & 654.7/659.7 & 1.1/2.3 & 355 & 0.197\\ 
\cutinhead{CDE Model}
15 & 654.9/660.6 & 2.1/7.7 & \nodata & \nodata \\
30 & 655.0/660.7 & 2.1/6.6 & \nodata & \nodata \\
45 & 655.0/660.7 & 1.9/5.3 & \nodata & \nodata \\
60 & 655.1/661.0 & 1.7/4.6 & \nodata & \nodata \\
75 & 655.3/661.0 & 1.6/4.3 & \nodata & \nodata \\
\cutinhead{Homogeneous Spheres Model}
15 & 655.6/662.6 & 1.4/3.6 & \nodata & \nodata \\
30 & 655.6/662.5 & 1.3/2.4 & \nodata & \nodata \\
45 & 655.6/662.4 & 1.1/0.9 & \nodata & \nodata \\
60 & 655.6/662.3 & 0.9/0.5 & \nodata & \nodata \\
75 & 655.8/662.4 & 0.6/0.3 & \nodata & \nodata \\
\cutinhead{Ice-coated Spheres Model\tablenotemark{b}}
15 & 655.0/660.4/666.4 & 2.0/13.7 & \nodata & \nodata \\
30 & 655.0/660.4/665.8 & 1.9/12.8 & \nodata & \nodata \\
45 & 655.0/660.4/664.8 & 1.7/\phantom{1}9.6 & \nodata & \nodata \\
60 & 655.1/660.5/664.4 & 1.4/\phantom{1}6.3 & \nodata & \nodata \\
75 & 655.2/660.6/664.4 & 0.7/\phantom{1}4.8 & \nodata & \nodata
\enddata
\tablecomments{The band position and \emph{FWHM} are listed for the multiple peaks within the CO$_{2}$ bending mode.}
\tablenotetext{a}{1 mono-layer (ML) $\approx$ 10$^{15}$ molecules cm$^{-2}$.}
\tablenotetext{b}{For broad profiles exhibiting blended peaks, the \emph{FWHM} value represents the overall width of the profile containing multiple peaks.}
\label{tbl:labpureco2}
\end{deluxetable}

Given the absorption coefficient, the real part of the complex refractive index, \emph{n}, is then calculated from the Kramers-Kronig dispersion relationship:

\begin{equation}
n(\tilde{\nu}) = n_{0} + \frac{1}{2 \pi^{2}} 
	\thinspace \mathscr{P}\negthinspace\int_{400}^{4000} \frac{\alpha(\tilde{\nu}^{\prime})}{\tilde{\nu}^{\prime2}-\tilde{\nu}^{2}} \, d\tilde{\nu}^{\prime},
\end{equation}

\begin{figure}[htp]
\centering
\rotatebox{90}{\includegraphics*[width=0.58\textwidth]{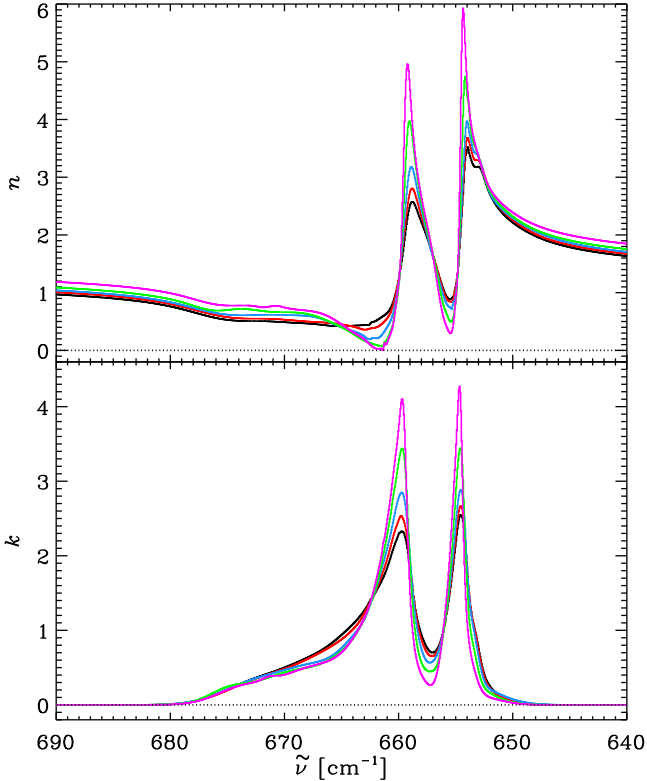}}
\caption{The optical constants (\emph{n} and \emph{k}) for the $\tilde{\nu}_{2}$ bending mode of pure CO$_{2}$ ice at laboratory temperatures of \emph{T} = 15 (black line), 30 (red line), 45 (blue line), 60 (green line), and 75 K (magenta line).}
\label{fig:optcontants}
\end{figure}

\noindent where $\mathscr{P}$ denotes the Cauchy principal value of the integral and $n_{0}$ is the refractive index for the high-wavenumber end of the infrared spectrum ($\tilde{\nu}$ $>$ 4000 cm$^{-1}$).  Because the visible ($\lambda = 0.633$ $\micron$) refractive index of pure CO$_{2}$ ice is temperature-dependent, we adopt values of $n_{0}$ $\approx$ 1.23, 1.26, 1.29, and 1.34 for the laboratory temperatures of \emph{T} = 15, 30, 45, and 60 K, respectively \citep{Satorre08}.  For the absorption spectrum at \emph{T} = 75 K, we use a value of $n_{0}$ $\approx$ 1.44 from \citet{Seiber71} to ensure that \emph{n} is positive for all wavenumbers.  Optical constants for the $\tilde{\nu}_{2}$ bending mode of pure CO$_{2}$ are presented in Figure \ref{fig:optcontants}.  

While our study is not intended to address the differences between the CO$_{2}$ optical constants derived in this work and those found among the literature, we do note, however, that any discrepancies are most likely due to differences in ice sample thickness and the $n_{0}$ refractive index for the high-wavenumber region ($\tilde{\nu}$ $>$ 4000 cm$^{-1}$).  For a thorough discussion on the discrepancies between CO$_{2}$ optical constants derived from different studies, we refer the reader to \citet{Ehrenfreund97}.        

\subsection{Grain Size and Shape Corrections}

\begin{figure}[htp]
\centering
\rotatebox{90}{\includegraphics*[width=0.679\textwidth]{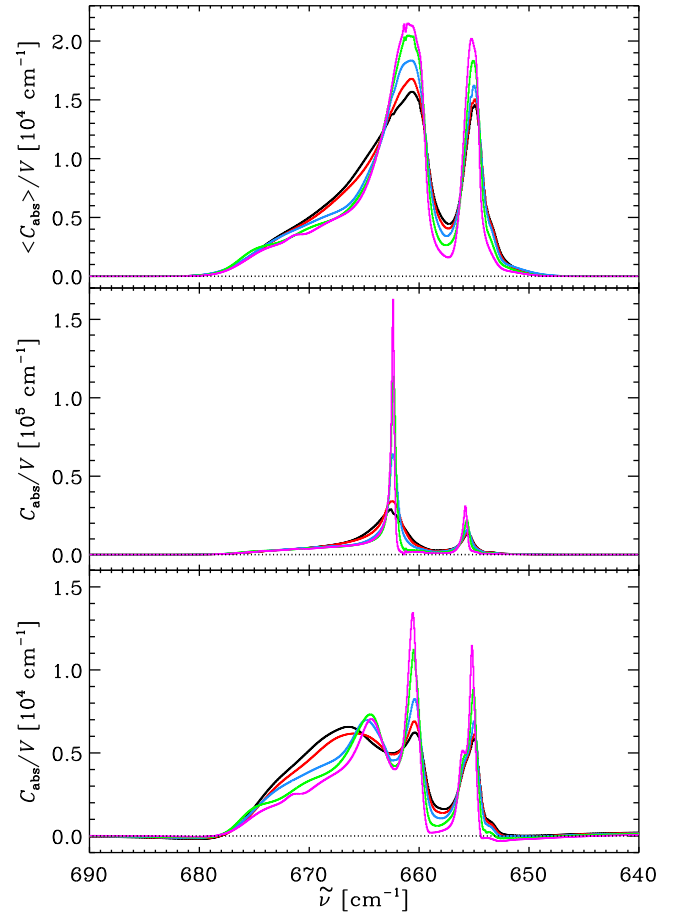}}
\caption{The absorption cross section per unit volume for the $\tilde{\nu}_{2}$ bending mode of pure CO$_{2}$ ice.  Grain shape models in the Rayleigh limit are shown for a continuous distribution of ellipsoids (CDE; top panel), homogeneous spheres (middle panel), and ice-coated silicate spheres (bottom panel) at laboratory temperatures of \emph{T} = 15 (black line), 30 (red line), 45 (blue line), 60 (green line), and 75 K (magenta line).}
\label{fig:grainshapes}
\end{figure}

The profiles of strong infrared resonance features, including those of ices, depend significantly on the shape and size of the absorbing and scattering particles \citep{Tielens91, Dartois06}.  For pure CO$_{2}$ ice, grain size and shape corrections must be applied to the laboratory absorption spectra of thin films to allow for an accurate comparison to astronomical spectra.  While irregularly-shaped grain models have been largely successful at simulating astronomical spectra of CO and CO$_{2}$ ices \citep{Pontoppidan03, Gerakines99}, we also consider here the shape effects of spherical grain models. 

Using the derived optical constants of pure CO$_{2}$, standard formulae from \citet{Bohren83} were applied in the Rayleigh limit to calculate the absorption cross section per unit volume,  \emph{C}$_{\rm{abs}}$/\emph{V}, for a continuous distribution of ellipsoids (CDE; all shapes are equally probable), homogeneous spheres, and ice-coated amorphous silicate spheres.  For the ice-coated spheres model, we adopt amorphous silicate optical constants from \citet{Draine84} and assume that half of the total particle volume is occupied by the core.  Note, however, for thick ice mantles or silicate cores that occupy $\lesssim$ 10\% of the total particle volume, the ice-coated spheres profile approximates that of pure homogeneous spheres.  

\begin{figure}[htp]
\centering
\rotatebox{90}{\includegraphics*[width=0.544\textwidth]{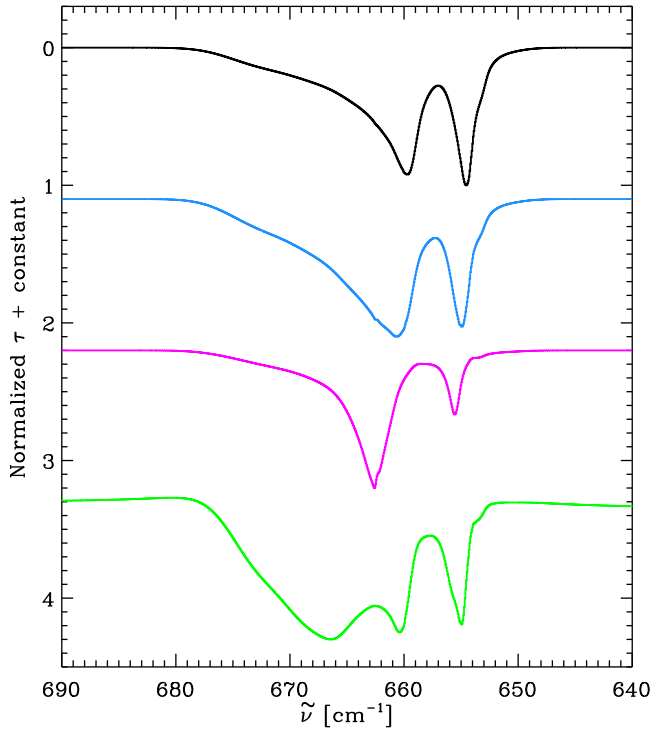}}
\caption{The normalized optical depth profiles for the $\tilde{\nu}_{2}$ bending mode of pure CO$_{2}$ ice. The uncorrected laboratory spectrum at \emph{T} = 15 K (black line) is compared to three standard grain shape models in the Rayleigh limit: a continuous distribution of ellipsoids (CDE; blue line), homogeneous spheres (magenta line), and ice-coated amorphous silicate spheres (green line).}
\label{fig:gshapesvslab}
\end{figure}

The absorption cross sections for the bending mode of CO$_{2}$ are presented in Figure \ref{fig:grainshapes} for the different grain shape models.  The CDE model produces a broad double-peaked absorption profile that remains mostly constant with increasing temperature from \emph{T} = 15 to 75 K; the peak positions exhibit a blueshift of 0.4 cm$^{-1}$ at \emph{T} = 75 K.  Conversely, a narrow double-peaked structure is generated by the homogenous spheres model, whose \emph{FWHM}s become narrower with increasing temperature;  the \emph{FWHM} of the 662.5 cm$^{-1}$ peak decreases by a factor of 12 from \emph{T} = 15 to 75 K.  In agreement with \citet{Ehrenfreund97}, we find that the ice-coated silicate spheres model gives rise to a broad triple-peaked profile.

A comparison between the different grain shape models and the laboratory spectrum of pure CO$_{2}$ at \emph{T} = 15 K is presented in Figure \ref{fig:gshapesvslab}.  For all models, the peak positions and \emph{FWHM}s deviate from those of the original laboratory spectrum.  For the CDE model, we find that the profile becomes blueshifted and broadened by $\sim$1 and $\sim$2.5 cm$^{-1}$, respectively.  For the homogeneous spheres model, a large blueshift of $\sim$3 cm$^{-1}$ and a narrowing of $\sim$2 cm$^{-1}$ are observed compared to the laboratory spectrum.  For a full comparison of the spectra, the reader is referred to Table \ref{tbl:labpureco2}.

{}


\begin{thebibliography}{}

\bibitem[An et al.(2011)]{An11} An, D., et al.\ 2011, \apj, 736, 133 

\bibitem[Andr{\'e} et al.(2010)]{Andre10} Andr{\'e}, P., et al.\ 2010, \aap, 518, L102

\bibitem[Arce et al.(2007)]{Arce07} Arce, H.~G., Shepherd, D., Gueth, F., Lee, C.-F., Bachiller, R., Rosen, A., Beuther, H.\ 2007, in Protostars and Planets V, Vol. 951, ed. B.~Reipurth, D.~Jewitt, \& K.~Keil (Tucson, AZ: Univ. Arizona Press), 245 

\bibitem[Baratta \& Palumbo(1998)]{Baratta98} Baratta, G.~A., \& Palumbo, M.~E.\ 1998, Journal of the Optical Society of America A, 15, 3076

\bibitem[Bergin et al.(1998)]{Bergin98} Bergin, E.~A., Neufeld, D.~A., \& Melnick, G.~J.\ 1998, \apj, 499, 777

\bibitem[Bergin et al.(1999)]{Bergin99} Bergin, E.~A., Neufeld, D.~A., \& Melnick, G.~J.\ 1999, \apj, 510, L145

\bibitem[Bergin et al.(2005)]{Bergin05} Bergin, E.~A., Melnick, G.~J., Gerakines, P.~A., Neufeld, D.~A., \& Whittet, D.~C.~B.\ 2005, \apj, 627, L33 

\bibitem[Billot et al.(2012)]{Billot12} Billot, N., Morales-Calder{\'o}n, M., Stauffer, J.~R., Megeath, S.~T., \& Whitney, B.\ 2012, \apj, 753, L35 

\bibitem[Bohren \& Huffman(1983)]{Bohren83} Bohren, C.~F., \& Huffman, D.~R.\ 1983, Absorption and Scattering of Light by Small Particles (New York: Wiley)

\bibitem[Boogert et al.(2002)]{Boogert02} Boogert, A.~C.~A., Blake, G.~A., \& Tielens, A.~G.~G.~M.\ 2002, \apj, 577, 271

\bibitem[Boogert et al.(2004)]{Boogert04} Boogert, A.~C.~A., et al.\ 2004, \apjs, 154, 359

\bibitem[Boogert et al.(2008)]{Boogert08} Boogert, A.~C.~A., et al.\ 2008, \apj, 678, 985

\bibitem[Bouwman et al.(2007)]{Bouwman07} Bouwman, J., Ludwig, W., Awad, Z., \"{O}berg, K. I., Fuchs, G. W., van Dishoeck, E.~F., \& Linnartz, H.\ 2007, \aap, 476, 995

\bibitem[Charnley(1997)]{Charnley97} Charnley, S.~B.\ 1997, \apj, 481, 396 

\bibitem[Cook et al.(2011)]{Cook11} Cook, A.~M., Whittet, D.~C.~B., Shenoy, S.~S., Gerakines, P.~A., White, D.W., \& Chiar, J.~E.\ 2011, \apj, 730, 124

\bibitem[Dartois et al.(1999)]{Dartois99} Dartois, E., Demyk, K., d'Hendecourt, L., \& Ehrenfreund, P.\ 1999, \aap, 351, 1066

\bibitem[Dartois(2006)]{Dartois06} Dartois, E.\ 2006, \aap, 445, 959

\bibitem[Davydov(1962)]{Davydov62} Davydov, A.~S.\ 1962, Theory of Molecular Excitons (McGraw-Hill: New York)

\bibitem[D'Hendecourt \& Jourdain de Muizon(1989)]{dHendecourt89} D'Hendecourt, L.~B., \& Jourdain de Muizon, M.\ 1989, \aap, 223, L5 

\bibitem[Draine \& Lee(1984)]{Draine84} Draine, B.~T., \& Lee, H.~M.\ 1984, \apj, 285, 89

\bibitem[Dunham \& Vorobyov(2012)]{Dunham12} Dunham, M.~M., \& Vorobyov, E.~I.\ 2012, \apj, 747, 52

\bibitem[Ehrenfreund et al.(1997)]{Ehrenfreund97} Ehrenfreund, P., Boogert, A.~C.~A., Gerakines, P.~A., Tielens, A.~G.~G.~M., \& van Dishoeck, E.~F.\ 1997, \aap, 328, 649 

\bibitem[Ehrenfreund et al.(1998)]{Ehrenfreund98} Ehrenfreund, P., Dartois, E., Demyk, K., \& d'Hendecourt, L.\ 1998, \aap, 339, L17

\bibitem[Ehrenfreund et al.(1999)]{Ehrenfreund99} Ehrenfreund, P., et al.\ 1999, \aap, 350, 240

\bibitem[Fischer et al.(2012)]{Fischer12} Fischer, W.~J., et al.\ 2012, \apj, 756, 99

\bibitem[Fraser et al.(2001)]{Fraser01} Fraser, H.~J., Collings, M.~P., McCoustra, M.~R.~S., \& Williams, D. A.\ 2001, \mnras, 327, 1165

\bibitem[Furlan et al.(2008)]{Furlan08} Furlan, E., et al.\ 2008, \apjs, 176, 184

\bibitem[Garrod \& Pauly(2011)]{Garrod11} Garrod, R.~T., \& Pauly, T.\ 2011, \apj, 735, 15

\bibitem[Gatley et al.(1974)]{Gatley74} Gatley, I., Becklin, E.~E., Mattews, K., Neugebauer, G., Penston, M.~V., \& Scoville, N.\ 1974, \apj, 191, L121

\bibitem[Gerakines et al.(1995)]{Gerakines95} Gerakines, P.~A., Schutte, W.~A., Greenberg, J.~M., \& van Dishoeck, E.~F.\ 1995, \aap, 296, 810

\bibitem[Gerakines et al.(1999)]{Gerakines99} Gerakines, P.~A., et al.\ 1999, \apj, 522, 357 

\bibitem[Hartmann et al.(1994)]{Hartmann94} Hartmann, L., Boss, A., Calvet, N., \& Whitney, B.\ 1994, \apj, 430, L49 

\bibitem[Hartmann et al.(1996)]{Hartmann96a} Hartmann, L., Calvet, N., \& Boss, A.\ 1996, \apj, 464, 387

\bibitem[Hartmann \& Kenyon(1996)]{Hartmann96b} Hartmann, L., \& Kenyon, S.~J.\ 1996, \araa, 34, 207 

\bibitem[Higdon et al.(2004)]{Higdon04} Higdon, S.~J.~U., et al.\ 2004, \pasp, 116, 975

\bibitem[Houck et al.(2004)]{Houck04} Houck, J., et al.\ 2004, \apjs, 154, 18

\bibitem[Hudgins et al.(1993)]{Hudgins93} Hudgins, D.~M., Sandford, S.~A., Allamandola, L.~J., \& Tielens, A.~G.~G.~M.\ 1993, \apjs, 86, 713

\bibitem[Ioppolo et al.(2011)]{Ioppolo11} Ioppolo, S., van Boheemen, Y., Cuppen, H.~M., van Dishoeck, E.~F., \& Linnartz, H.\ 2011, \mnras, 413, 2281

\bibitem[Keesom \& K\"{o}hler(1934)]{Keesom34} Keesom, W.~H., \& K\"{o}hler, J.~W.~L.\ 1934, Physica, 1, 655

\bibitem[Kim et al.(2012)]{Kim12} Kim, H.~J., Evans, N.~J., II, Dunham, M.~M., Lee, J.-E., \& Pontoppidan, K.~M.\ 2012, arXiv:1208.5797 

\bibitem[Knez et al.(2005)]{Knez05} Knez, C., et al.\ 2005, \apj, 635, L145

\bibitem[Kryukova et al.(2012)]{Kryukova12} Kryukova, E., Megeath, S.~T., Gutermuth, R.~A., Pipher, J., Allen, T.~S., Allen, L.~E., Myers, P.~C., \& Muzerolle, J.\ 2012, \aj, 144, 31 

\bibitem[Markwardt(2009)]{Markwardt09} Markwardt, C.~B.\ 2009, in ASP Conf. Ser. 411, Astronomical Data Analysis Software and Systems XVIII, ed. D.~A. Bohlender, D. Durand, \& P. Dowler (San Francisco, CA: ASP), 251

\bibitem[Mastrapa et al.(2009)]{Mastrapa09} Mastrapa, R.~M., Sandford, S.~A., Roush, T.~L., Cruikshank, D.~P., \& Dalle Ore, C.~M.\ 2009, \apj, 701, 1347

\bibitem[McKee \& Ostriker(2007)]{McKee07} McKee, C.~F., \& Ostriker, E.~C.\ 2007, \araa, 45, 565 

\bibitem[Mennella et al.(2004)]{Mennella04} Mennella, V., Palumbo, M.~E., \& Baratta, G.~A.\ 2004, \apj, 615, 1073 

\bibitem[Menten et al.(2007)]{Menten07} Menten, K.~M., Reid, M.~J., Forbrich, J., \& Brunthaler, A.\ 2007, \aap, 474, 515

\bibitem[Mezger et al.(1990)]{Mezger90} Mezger, P.~G., Zylka, R., \& Wink, J.~E.\ 1990, \aap, 228, 95

\bibitem[Molinari et al.(2010)]{Molinari10} Molinari, S., et al.\ 2010, \aap, 518, L100

\bibitem[Neufeld \& Hollenbach(1994)]{Neufeld94} Neufeld, D.~A., \& Hollenbach, D.~J.\ 1994, \apj, 428, 170

\bibitem[Noble et al.(2011)]{Noble11} Noble, J.~A., Dulieu, F., Congiu, E., \& Fraser, H. J.\ 2011, \apj, 735, 121

\bibitem[Noble et al.(2012)]{Noble12} Noble, J.~A., Congiu, E., Dulieu, F., \& Fraser, H.~J.\ 2012, \mnras, 421, 768

\bibitem[Nummelin et al.(2001)]{Nummelin01} Nummelin, A., Whittet, D.~C.~B., Gibb, E.~L., Gerakines, P.~A., \& Chiar, J.~E.\ 2001, \apj, 558, 185 

\bibitem[{\"O}berg et al.(2009)]{Oberg09} {\"O}berg, K.~I., Fayolle, E.~C., Cuppen, H.~M., van Dishoeck, E.~F., \& Linnartz, H.\ 2009, \aap, 505, 183 

\bibitem[{\"O}berg et al.(2011)]{Oberg11} {\"O}berg, K.~I., Boogert, A.~C.~A., Pontoppidan, K.~M., van den Broek, S., van Dishoeck, E.~F., Bottinelli, S., Blake, G.~A., Evans, N.~J., II.\ 2011, \apj, 740, 109

\bibitem[Oba et al.(2010)]{Oba10} Oba, Y., Watanabe, N., Kouchi, A., Hama, T., \& Pirronello, V.\ 2010, \apj, 712, L174 

\bibitem[Oliveira et al.(2009)]{Oliveira09} Oliveira, J.~M., et al.\ 2009, \apj, 707, 1269

\bibitem[Ormel et al.(2009)]{Ormel09} Ormel, C.~W., Paszun, D., Dominik, C., \& Tielens, A.~G.~G.~M.\ 2009, \aap, 502, 845

\bibitem[Ossenkopf(1993)]{Ossenkopf93} Ossenkopf, V.\ 1993, \aap, 280, 617

\bibitem[Pontoppidan et al.(2003)]{Pontoppidan03} Pontoppidan, K.~M., et al.\ 2003, \aap, 408, 981

\bibitem[Pontoppidan et al.(2008)]{Pontoppidan08} Pontoppidan, K.~M., et al.\ 2008, \apj, 678, 1005

\bibitem[Poteet et al.(2011)]{Poteet11} Poteet, C.~A., et al.\ 2011, \apj, 733, L32

\bibitem[Reipurth \& Aspin(2010)]{Reipurth10} Reipurth, B., \& Aspin, C.\ 2010, in Evolution of Cosmic Objects through their Physical Activity, ed. H.~A.~Harutyunian, A.~M.~Mickaelian, \& Y.~Terzian (Yerevan: Gitutyum), 19

\bibitem[Roser et al.(2001)]{Roser01} Roser, J.~E., Vidali, G., Manic{\`o}, G., \& Pirronello, V.\ 2001, \apj, 555, L61 

\bibitem[Ruffle \& Herbst(2001)]{Ruffle01} Ruffle, D.~P., \& Herbst, E.\ 2001, \mnras, 324, 1054 

\bibitem[Satorre et al.(2008)]{Satorre08} Satorre, M.~{\'A}., Domingo, M., Mill{\'a}n, C., et al.\ 2008, \planss, 56, 1748

\bibitem[Seale et al.(2011)]{Seale11} Seale, J.~P., Looney, L.~W., Chen, C.-H.~R., Chu, Y.-H., \& Gruendl, R.~A.\ 2011, \apj, 727, 36

\bibitem[Seiber et al.(1971)]{Seiber71} Seiber, B.~A., Smith, A.~M., Wood, B.~E., \& M\"{u}ller, P.~R.\ 1971, \ao, 10, 2086

\bibitem[Terebey et al.(1984)]{Terebey84} Terebey, S., Shu, F.~H., \& Cassen, P.\ 1984, \apj, 286, 529

\bibitem[Tielens \& Hagen(1982)]{Tielens82} Tielens, A.~G.~G.~M., \& Hagen, W.\ 1982, \aap, 114, 245

\bibitem[Tielens(1991)]{Tielens91} Tielens, A.~G.~G.~M., Tokunaga, A.~T., Geballe, T~ R., \& Baas, F.\ 1991, \apj, 381, 181

\bibitem[Tobin et al.(2010)]{Tobin10} Tobin, J.~J., Hartmann, 
L., Looney, L.~W., \& Chiang, H.-F.\ 2010, \apj, 712, 1010 

\bibitem[Tobin et al.(2011)]{Tobin11} Tobin, J.~J., et al.\ 2011, \apj, 740, 45 

\bibitem[Tobin et al.(2012)]{Tobin12} Tobin, J.~J., Hartmann, L., Bergin, E., Chiang, H.-F., Looney, L.~W., Chandler, C.~J., Maret, S., \& Heitsch, F.\ 2012, \apj, 748, 16

\bibitem[van Broekhuizen et al.(2006)]{vanBroekhuizen06} van Broekhuizen, F.~A., Groot, I.~M.~N., Fraser, H.~J., van Dishoeck, E.~F., \& Schlemmer, S.\ 2006, \aap, 451, 723 

\bibitem[Werner et al.(2004)]{Werner04} Werner M., et al.\ 2004, \apjs, 154, 1

\bibitem[White et al.(2009)]{White09} White, D.~W., Gerakines, P.~A., Cook, A.~M., \& Whittet, D.~C.~B.\ 2009, \apjs, 180, 182

\bibitem[Whittet et al.(1998)]{Whittet98} Whittet, D.~C.~B., Gerakines, P.~A., Tielens, A.~G.~G.~M., et al.\ 1998, \apj, 498, L159

\bibitem[Whittet et al.(2007)]{Whittet07} Whittet, D.~C.~B., Shenoy, S.~S., Bergin, E.~A., Chiar, J.~E., Gerakines, P.~A., Gibb, E.~L., Melnick, G.~L., \& Neufeld, D.~A.\ 2007, \apj, 655, 332

\bibitem[Whittet et al.(2009)]{Whittet09} Whittet, D.~C.~B., Cook, A.~M., Chiar, J.~E., Pendleton, Y.~J., Shenoy, S.~.S., \& Gerakines, P.~A.\ 2009, \apj, 695, 94
 
\bibitem[Williams et al.(2003)]{Williams03} Williams, J.~P., Plambeck, R.~L., \& Heyer, M.~H.\ 2003, \apj, 591, 1025

\bibitem[Zasowski et al.(2009)]{Zasowski09} Zasowski, G., Kemper, F., Watson, D.~M., Furlan, E., Bohac, C.~J, Hull, C., \& Green, J.~D.\ 2009, \apj, 694, 459

\end{thebibliography}
\end{document}